\documentstyle[aps,eqsecnum,epsf]{revtex}

\newcommand{\A}{{\mathbf A}}
\newcommand{\B}{{\mathbf B}}
\newcommand{\cc}{\mathrm{c.c.}}
\newcommand{\C}{{\mathbf C}}
\renewcommand{\d}{{\mathrm d}}
\newcommand{\D}{{\mathbf D}}
\newcommand{\e}{{\mathrm e}}
\newcommand{\eq}{&=&}

\newcommand{\g}[1]{\gamma_{#1}}
\newcommand{\G}{{\mathbf G}}
\renewcommand{\i}{{\mathbf i}}
\newcommand{\intz}{\int\!\d z_i\,}
\newcommand{\J}{{\mathbf J}}
\newcommand{\K}{{\mathbf K}}
\newcommand{\M}{{\mathbf M}}
\renewcommand{\O}{{\mathbf O}}
\renewcommand{\P}{{\mathcal P}}
\newcommand{\ps}{\overline{\psi}}
\newcommand{\prodn}[1]{\prod_{i=#1}^n}
\newcommand{\Q}{{\mathcal Q}}

\newcommand{\R}{{\mathcal R}}
\newcommand{\T}{T}
\newcommand{\Tb}{\overline{\T}}
\newcommand{\Tm}{{\mathbf T}}
\newcommand{\myfigure}[3]{\begin{figure}%
	\epsfxsize#1%
	\centerline{\hspace{0.15in}\epsffile{#2.eps}}%
	\caption{#3}\label{fig:#2}\end{figure}}

\begin{document}

\twocolumn[\hsize\textwidth\columnwidth\hsize\csname@twocolumnfalse\endcsname
\title{Effective attraction between like-charged colloids in a 2D plasma}
\author{Ning Ma$^{a}$, S. M. Girvin$^{a}$ and R. Rajaraman$^{b}$}
\address{(a)Department of Physics, Indiana University, 
Bloomington, IN 47405 USA\\
(b)School of Physical Sciences, Jawaharlal Nehru University, 
New Delhi, 110067 India}
\maketitle
\begin{abstract}
{The existence of attractions between like-charged colloids
immersed in ionic solution have been discovered in recent
experiments.  This phenomenon contradicts the predictions of
DLVO theory and indicates a failure of mean field theory. 
We study a toy model based on a two 
dimensional one-component plasma, which is exactly soluble
at one particular coupling constant.  We show that colloidal
interaction results from a competition between ion-ion repulsion
and longer ranged ion-void attraction.}{}
\end{abstract}

\vskip2pc]

\section{Introduction} 
Recent experiments show convincingly the existence of attractions between 
like-charged colloids immersed in ionic solution, in particular
in the vicinity of a glass wall or when the colloids are
confined between glass walls \cite{Kepler,Grier}. 
This remarkable counter-intuitive phenomenon is
inconsistent with the well established theory of Derjaguin,
Landau, Verwey and Overbeek (DLVO) and has generated various theoretical
interpretations.  

It was shown \cite{Parsegian} that counterion
correlation forces, because of the long range of the
electrostatic potentials, are usually not pairwise additive. 
Many-body effects \cite{Schmitz} and Coulomb depletion forces 
\cite{Allahyarov} have been invoked to explain attractive
forces between like-charged macroions.
An exact demonstration \cite{Neu} proved that the non-linear 
Poisson-Boltzmann (PB) mean-field equation
can not give attractions in the case of Dirichlet boundary
conditions.  The argument was extended to a broader class of
models \cite{Raimbault} and boundary conditions \cite{Sader}.
Recently, there was a significant new attempt based on 
non-equilibrium hydrodynamic effects \cite{bren}.

In this paper, we solve a 2D one-component plasma model exactly
at a certain coupling constant following the method of 
Jancovici \cite{Jaco}.  We have found a semi-analytic method which 
extends the exact solution to the case in which colloidal particles
are present.  We treat the colloids in the plasma as empty 
regions (voids) from which the ions  are excluded.
The colloid may have fixed charge at its center. 
The analytic results obtained for small-sized colloids 
show that while the ion-ion repulsion, which goes like $\e^{-R^2}$
($R$ is the separation distance)
in this particular 2D model, is strong at short distances, the
ion-void attraction, which goes like $R^2\e^{-R^2}$, becomes
dominant at large distances.  The latter property is not
normally found in conventional mean field approximations, and
must come from strong correlation effects.
For large-sized colloids, we also have numerical results which
qualitatively agree with the analytic ones.  
The present state of understanding of how to treat reliably
fluctuations beyond mean field theory is not very good despite
the fact that this is an old problem \cite{rich,orland,brown}.
Thus, our exact results
may serve as a testing ground for more sophisticated
approximation schemes.  
In addition, since this is a 2D model, the results are of
interest in connection with the problem of interactions among
linear polyelectrolyte molecules \cite{ajl}.

In the sections that follow, we shall first review the plasma model in
section \ref{chap:p}.  Then in section \ref{chap:i} and 
\ref{chap:v} we calculate the ion-ion 
and void-void  interactions in the plasma system respectively.  Section
\ref{chap:c} puts the above results together by
treating the colloid as an empty region plus a fixed ion charge at the center.
We have relegated the technical details as much as
possible to the appendices.

\section{The Plasma System}\label{chap:p}
We start by reviewing the exact solution for the one-component 2D
plasma system \cite{Jaco}. Our approach is to map the classical plasma
system into a quantum Hall system.   Such an analogy allows one 
to describe the interacting ions in terms of non-interacting 
electrons occupying certain quantum obitals, and
subsequently use already developed tools.
The advantage of this approach will become more clear later 
when we extend the same analogy to more complicated systems 
with fixed ion charges or/and empty regions present.

\subsection{The Quantum Orbital Language}\label{sec:pm}
The one-component 2D plasma system is composed of $n$ simple 
ions embedded in a disk of uniform background charge.
Each ion carries a charge of $-1$; the disk has a radius $r$ and 
an areal charge density of $\rho_0$.  Throughout this paper, we
shall express all lengths in the unit of
$\ell_0={1\over\sqrt{\pi\rho_0}}$.  Thus, a unit circle 
contains unit amount of background charge, and the charge neutrality
condition can be written simply as \begin{equation}
	 n=r^2.
\end{equation}
In two dimensions, it is convenient to use a complex number 
$z=x+\i y$ (in units of $\ell_0$) to represent the location of 
$(x,y)$.  Using $z_i$ to denote the position of the $i$th ion, we can 
solve the 2D Poisson equation
and find the total Coulomb potential energy of the system:
\begin{equation}
	 V={1\over 2}\sum_{i=1}^n|z_i|^2 -\sum_{i<j}^n\ln|z_i-z_j|.
\end{equation}
Here the first term describes the attractions between the ions and the 
background charge, and the second term represents the mutual repulsions
among the ions themselves.  

At inverse temperature $\beta=2$, the Boltzmann factor reads:
\begin{equation}
	 \e^{-\beta V}=\e^{-\sum_{i=1}^n |z_i|^2} 
	 	\left|\prod_{i<j}^n (z_i-z_j)\right|^2.\label{pm1}
\end{equation}
Written in the above form, the Boltzmann factor can be identified
as the square of the determinant of a matrix $\M$ 
\begin{equation}
	\e^{-\beta V}=|\det \M|^2, \label{pm2}
\end{equation}
with the matrix elements being the lowest Landau level (LLL)
wavefunctions studied in connection with the quantum Hall effect
\cite{smg}: \begin{equation}
	 M_{ij}=\psi_i(z_j)\equiv{z_j^{i-1}\over \sqrt{\pi \Gamma(i)}}
	 \e^{-|z_j|^2/2}.\label{pm3}
\end{equation}
The above wavefunction describes an electron occupying the 
$i$th angular momentum orbital in the LLL.
In this quantum language, $\det \M$ is a Slater determinant and
represents the Laughlin wavefunction at filling fact $\nu=1$,
describing a fully filled Landau level.  Thus, the complicated
Coulomb interactions are replaced by a simple Slater
determinant.  The determinant
representation of the Boltzmann factor was first noted by
Jancovici \cite{Jaco} prior to the discovery of the quantum Hall
effect.  We recapitulate (and then extend) his argument in more
modern language.

\subsection{The Partition Function}\label{sec:pp}
Armed with this powerful quantum analogy, we proceed to calculate
the partition function of the system, by
integrating out all the ionic degrees of freedom:
\begin{equation}
    Z=\prodn{1}\intz\e^{-\beta V}. \label{pp1}
\end{equation}
This is nothing more than the norm of the Laughlin wavefunction,
and hence can be computed exactly.
Substituting eq.(\ref{pm2}) in eq.(\ref{pp1}) 
and expanding the determinant, we have: 
\begin{eqnarray}
	 Z\eq\prodn{1}\intz|\det \M|^2\nonumber\\
	 \eq\prodn{1}\intz
	 	\sum_{\{\P\}}\sum_{\{\Q\}}(-1)^{\P+\Q}\prodn{1}
		\ps_{P_i}(z_i)\psi_{Q_i}(z_i),\label{pp2}
\end{eqnarray}
where $\P$ and $\Q$ are permutations of
$\{1:n\}$ [$1:n$ is a short hand notation for
$1,2,\ldots,n$] that come 
from the expansions of $\det\overline{\M}$ and $\det \M$ respectively.
In eq.(\ref{pp2}), we have both summations and integrations which, 
respectively represent the Coulomb interactions and the thermal
averages.  It is useful at this point to postpone
treating the Coulomb interactions until after the thermal
averages are done.  Thus, we switch the order of the summations and 
integrations in eq.(\ref{pp2}) so that: \begin{eqnarray}
	 Z\eq\sum_{\{\P\}}\sum_{\{\Q\}}(-1)^{\P+\Q}
		 \prodn{1}\intz \ps_{P_i}(z_i)\psi_{Q_i}(z_i)
		 \nonumber\\
	 &\equiv& \sum_{\{\P\}}\sum_{\{\Q\}}(-1)^{\P+\Q}
	 	\prodn{1}\langle P_i|Q_i\rangle, \label{pp3}
\end{eqnarray}
where $\langle i|j\rangle$ is the Dirac notation for the inner product 
of the $i$th and $j$th LLL orbitals.
Notice that the summations over $\{\Q\}$ 
can be identified as a matrix determinant:
\begin{equation}
	 \sum_{\{\Q\}}(-1)^{\P+\Q}\prodn{1}\langle P_i|Q_i\rangle=
	 \sum_{\{\R\}}(-1)^{\R}\prodn{1}\langle i|R_i\rangle=\det\O,
	 \label{pp4}
\end{equation}
where $\R\equiv\Q\circ\P^{-1}$ is the composite permutation 
of $\Q$ and $\P^{-1}$,
and the {\it overlap matrix} $\O$ defined by
\begin{equation}
	 O_{ij}=\langle i|j\rangle\label{pp9}
\end{equation}
is a matrix composed of LLL wavefunction inner products. 
Here, just as previously, 
the complicated Coulomb interactions are again replaced by
a simple matrix determinant.
Finally, we substitute eq.(\ref{pp4}) into (\ref{pp3}) to
obtain:\begin{equation}
	Z=\sum_{\{\P\}}\det\O=n!\det\O.\label{pp5}
\end{equation}
The immaterial factor of $n!$ will be dropped henceforth.  
Through the quantum analogy, the statistical
problem is reduced to evaluating the overlap matrix $\O$ and
its determinant.  This is particularly simple in the present 
case,  because of orthonormality we know that matrix
$\O$ is an identity matrix.
\subsection{The Plasma Edge} \label{sec:pb}
Near the plasma edge,  physical properties 
are quite different from those in the bulk.   This motivates us
to study the boundary effects separately.
For simplicity, we are going to be mainly interested 
in a {\it soft} boundary condition,  
which restricts the range of the ions' angular
momenta rather than their positions.
The ions can go anywhere in the complex plane,
nevertheless, they are most likely to be found inside a disk
with radius $r=\sqrt{n}$ because their angular momenta are bounded
by $n$ [Recall that the $l$th LLL state has
angular momentum of $(l-1)$ and
peaks in a shell $\sqrt{l-1}<r<\sqrt{l}$].
The advantage of using this type of boundary condition is apparent: 
we can continue to use the same
wavefunctions to describe the ions near the edge.
What is implicitly assumed here is that the uniform background charge 
does not cease at the plasma edge; it extends to infinity.
We also note that in the thermodynamic limit ($r\to\infty$),
the circular plasma edge becomes locally flat.

One may also consider putting some `surface' charge on the plasma 
boundary. But, as indicated below, this effect can be
absorbed in shifting the position of the boundary.
We assume the surface charge, if any, does not fluctuate.  
Depending on its sign, the boundary charge either increases or
decreases the number of available angular momentum channels. 
This is more or less equivalent to varying the radius $r$ of the
plasma disk in a system with neutral boundary.
Hence, without loss of generality, we may assume boundaries are
neutral.

Another type of plasma system we want to address in this paper is a plasma 
strip with two parallel edges separated by $h$.
This can be realized in the thermodynamic limit by a system of
ions with angular momenta ranging from $(r-{h\over 2})^2$ to 
$(r+{h\over 2})^2$.
The double edged system thus differs from single edged system 
{\it only} in the momenta range.
To avoid redundancy, we shall use the single edged system as
example in our calculations.
By changing the angular momentum limits, the results can be easily
transferred to the double edged systems.

\section{Fixed Ion Charges in a Plasma}\label{chap:i}
In this section, we shall apply the same method to a slightly 
more complicated plasma system, with one or two ion charge(s) 
fixed at certain position(s).  
The resulting partition function is conventionally called the
one- or two-body density function.  The physical significance
is two-fold.  First, we can learn from these density functions 
about how the ions
are distributed and how two ions are correlated.  More
importantly, the fixed ion charges can be viewed as the limiting
case of small-sized charged colloids.  Thus, the results obtained 
in this section will provide us with many useful insights to the
colloidal interactions.

\subsection{The One-body Density Function}\label{sec:i1}
First  we consider the one-body density function.
We fix the position $z_1=w$
and integrate out the remaining ionic degrees of freedom
\begin{eqnarray}
	Z^{(1)}\eq\prodn{2}
		\left.\intz\e^{-\beta V}\right|_{z_1=w}\nonumber\\
	\eq\prodn{2}
		\left.\intz|\det\M|^2\right|_{z_1=w}\nonumber\\
	\eq\sum_{\{\P\}}\sum_{\{\Q\}}(-1)^{\P+\Q}
		\ps_{P_1}(w)\psi_{Q_1}(w)\prodn{2}\langle
		P_i|Q_i\rangle.\label{i10}
\end{eqnarray}
The summations over $P_1$ and $\{\Q\}$ then reduce to a matrix
determinant:
\begin{equation}
	\sum_{P_1}\sum_{\{\Q\}}(-1)^{\P+\Q}
		\ps_{P_1}(w)\psi_{Q_1}(w)\prodn{2}\langle
		P_i|Q_i\rangle=-\det\O^{(1)},
\end{equation}
with the $(n+1)$ by $(n+1)$ matrix $\O^{(1)}$ defined by:
\begin{equation}
	 \O^{(1)}\equiv 
	 \left[\begin{array}{cc}
	 	\O & \Psi^\dagger(w)\\
	 	\Psi(w) & 0
	\end{array}\right].\label{i11}
\end{equation}
Here, $\O$ (not to be confused with $\O^{(1)}$) is the $n\times
n$ overlap matrix defined in
(\ref{pp9}); $\Psi\equiv(\psi_1,\psi_2,\ldots,\psi_n)$ is a row
vector of LLL wavefunctions representing the fixed ion charge,  and
$\Psi^\dagger$ is its conjugate transpose.
Finally, the remaining summations over $\{P_{i,(i=2:n)}\}$ in eq.(\ref{i10}) 
just yield a trivial prefactor:
\begin{equation}
	 Z^{(1)}=\sum_{\{P_{i(i=2:n)}\}}-\det\O^{(1)}=-(n-1)!\det\O^{(1)}.
\end{equation}

To evaluate the determinant of matrix $\O^{(1)}$, we apply
Schur's theorem (discussed in detail in appendix
\ref{chap:ast}) and find: \begin{eqnarray}
	 Z^{(1)}\eq-\det\O^{(1)}=\det\O\cdot\det\left[
	 	\Psi(w)\O^{-1}\Psi^{\dagger}(w)\right]_{1\times1}\nonumber\\
	\eq\sum_{k=1}^n\psi_k(w)\ps_k(w),
\end{eqnarray}
where we have used the fact that the overlap matrix $\O$ is an identity
matrix.  Using eq.(\ref{atc0}), we convert the above series
into translation coefficients (TCs) defined in eq.(\ref{aoc4}):
\begin{equation}
	 Z^{(1)}=\sum_{k=1}^n\psi_k(w)\ps_k(w)
	 	={1\over\pi}\sum_{k=1}^n|\T_{k1}(w)|^2.
		\label{i15}
\end{equation}
Now that the partition function is written in the form of a TC sum,
we may use the TC sum-rules listed in appendix \ref{sec:atc}.
We consider the following two cases.

If the fixed ion is in the bulk of the plasma ($w\ll r=\sqrt{n}$), 
we may take the limit $n\to\infty$ first and use TC sum-rule (\ref{atc7})
to obtain:\begin{equation}
	 Z^{(1)}={1\over\pi}.
\end{equation}
This tells that deep in the bulk, the ion density is constant 
and equal to the background charge density everywhere.

Near the plasma edge, we assume the boundary is located at
$(r+{h\over 2})$ and  the fixed ion is located at 
$w=r\e^{\i\theta}$, so that the ion is ${h\over2}$ away from the
boundary.  In this case, we can use TC sum-rule (\ref{atc11}) in
eq.(\ref{i15}) to obtain: \begin{equation}
	 Z^{(1)s}={1\over 2\pi}\left[1+\Phi\left(h\over\sqrt{2}\right)\right].
\end{equation}
Here, the superscript `s' means single edged system (in contrast to `d', 
which means double edged system) and $\Phi(x)$ is the standard error
function. 
As shown in Fig.\ \ref{fig:f1}, the ion density indeed drops
from its bulk value to zero upon crossing the soft plasma
boundary (at $h=0$).

\subsection{The Two-body Density Function}\label{sec:i2}
In calculating the one-body density function, we find that
a fixed ion charge amounts to an additional row and column 
to the overlap matrix.  This result can be generalized for
systems with multiple fixed ion charges. 
In particular, we consider the two-body density function
with $z_1=w$ and $z_2=\overline{w}$ fixed.  The result 
is: \begin{equation}
	 Z^{(2)}=-\det\O^{(2)},\label{i20}
\end{equation}
where $\O^{(2)}$ is an $(n+2)$ by $(n+2)$ matrix:
\begin{equation}
	 \O^{(2)}\equiv \left[
	 \begin{array}{ccc}
	 	\O&\Psi^\dagger(w)&\Psi^\dagger(\overline{w})\\
		\Psi(w)&0&0\\
		\Psi(\overline{w})&0&0
	\end{array}
	 \right].\label{i21}
\end{equation}
Similarly using Schur's theorem and eq.(\ref{atc0}), 
we can write the correlation function in terms of TC sums:
\begin{eqnarray}
	 Z^{(2)}\eq\det\O\cdot\det\left(
	 	\left[\begin{array}{c}
			\Psi(w)	\\\Psi(\overline{w})
		\end{array}\right]\O^{-1}\left[\begin{array}{cc}
			\Psi^{\dagger}(w)&\Psi^{\dagger}(\overline{w})
		\end{array}\right]
		 \right)_{2\times 2}\nonumber\\
	 \eq{1\over\pi^2}\left[\sum_{k=1}^n|\T_{k1}(w)|^2\right]^2
		-{1\over\pi^2}\left|\sum_{k=1}^n\T_{k1}^2(w)\right|^2.
	\label{i25}
\end{eqnarray}

In the bulk of the plasma ($w\ll r$), we assume
$w=\i{R\over2}$ so that the two fixed ions are $R$ apart. 
Plugging TC sum-rules (\ref{atc7}) and (\ref{atc8}) into 
eq.(\ref{i25}), and using eq.(\ref{atc2}) for $\T_{11}(\i R)$, 
we obtain:
\begin{equation}
	 Z^{(2)}={1\over\pi^2}\left[1-|\T_{11}(\i R)|^2\right]
	 ={1\over\pi^2}\left(1-\e^{-R^2}\right).
	 \label{i26}
\end{equation}
The above result tells that the Coulomb repulsion between the
two ion charges is heavily screened by the other ions
and becomes short ranged.  Its decay has a Gaussian form, 
which is much faster than the exponential form that DLVO 
theory predicts.

Near the plasma edge, we assume the boundary is located at
$(r+{h\over2})$ and the fixed ions are located at 
$w=r\e^{\i\theta}$ and $\overline{w}$, where
$\theta\equiv{R\over 2r}$ so that the fixed ions are ${h\over2}$
away from the edge and $R$ apart from each other.  Using TC
sum-rules (\ref{atc11}) and (\ref{atc14}) in eq.(\ref{i25}), 
we obtain (for asymptotically large $R$): \begin{equation}
	Z^{(2)s}={1\over4\pi^2}\left\{\left[1+\Phi\left({h\over\sqrt{2}}\right)
		\right]^2-{2\e^{-h^2}\over\pi R^2}\right\}.
\end{equation}
The above result shows that the ion-ion correlation function
has a power law decay near the plasma edge.
This is fundamentally different from the
bulk behavior, and is due to the dipole moments
induced by the sharp cutoff in angular momentum.

Finally, in the double edged system where the plasma edges are
located at $(r\pm{h\over2})$, we
may use TC sum-rules (\ref{atc21}) and (\ref{atc24}) in eq.(\ref{i25})
to obtain (for asymptotically large $R$):\begin{equation}
	Z^{(2)d}={1\over\pi^2}\left[\Phi^2\left(h\over\sqrt{2}\right)
		-{\e^{-h^2}\over 2\pi R^2}\sin^2(hR)\right].
\end{equation}
Here, we find the correlation is not only long ranged, but
oscillating with a period of $2\pi h^{-1}$ as well.

\subsection{Divalent Ions} \label{sec:i4}
A divalent ion can be viewed as composed of two simple ions
occupying the same place.  Here we want to calculate the two-body 
density function for a pair of divalent ions embedded in the
monovalent plasma.
Let us assume $z_1=z_2=w$ and $z_3=z_4=\overline{w}$ are fixed,
where $w=\i{R\over2}$ so that the two divalent ions are $R$ apart.
Naively one may follow eq.(\ref{i21}) and
write down an $(n+4)$ by $(n+4)$ matrix:
\begin{equation}
	 \tilde\O^{(4)}\equiv \left[
	 \begin{array}{ccccc}
	 	\O&\Psi^\dagger(z_1)&\Psi^\dagger(z_2)
			&\Psi^\dagger(z_3)&\Psi^\dagger(z_4)\\
	 	\Psi(z_1)&0&0&0&0\\
	 	\Psi(z_2)&0&0&0&0\\
	 	\Psi(z_3)&0&0&0&0\\
	 	\Psi(z_4)&0&0&0&0
	\end{array}
	\right].\label{i41}
\end{equation}
However, the matrix $\tilde\O^{(4)}$ defined above is singular 
because $z_1=z_2$ (or $z_3=z_4$). 
This singularity actually originates from the Boltzman factor
in eq.(\ref{pm1}), where we incorrectly include the self interaction
between $z_1$ and $z_2$ (or $z_3$ and $z_4$).
To get around this problem, we must extract the singular factors from
the matrix determinant.  More precisely, we want to consider the
following limits:
\begin{equation}
	\det\O^{(4)}\equiv\lim_{z_2\to z_1}\lim_{z_4\to z_3} 
	{\det\tilde\O^{(4)}(z_1,z_2,z_3,z_4)\over|z_2-z_1|^2|z_4-z_3|^2}.
\end{equation}
To construct matrix $\O^{(4)}$ from $\tilde\O^{(4)}$,
we subtract $\Psi(z_1)$ in
the $(n+1)$th row from $\Psi(z_2)$ in the $(n+2)$th row
in eq.(\ref{i41}).  
The resulting $(n+2)$th row then yields a factor of $(z_2-z_1)$:
\begin{equation}
	 \lim_{z_2\to z_1}[\Psi(z_2)-\Psi(z_1)]=
	 (z_2-z_1)\Psi'(z_1),\label{dif}
\end{equation}
where $\Psi'\equiv(\psi_1',\psi_2',\ldots,\psi_n')$, and $\psi'$ is the 
first derivative of $\psi$.  More precisely, following the usual
procedure for the Hilbert space of analytic
functions \cite{smg,smg2}, we take the derivative to act
only on the analytic part of $\psi$ and not on the Gaussian
factor.  The latter represents the interaction with the
background charge and does not possess the self-interaction
singularity.

The factors of $(\overline{z}_2-\overline{z}_1)$ and
$|z_4-z_3|^2$ can be extracted via similar procedures and we are
left with a matrix that looks like the following:
\begin{equation}
	 \O^{(4)}\equiv\left[
	 \begin{array}{ccccc}
	 	\O&\Psi^\dagger(w)&{\Psi^\dagger}'(w)
			&\Psi^\dagger(\overline{w})&{\Psi^\dagger}'(\overline{w})\\
	 	\Psi(w)&0&0&0&0\\
	 	\Psi'(w)&0&0&0&0\\
	 	\Psi(\overline{w})&0&0&0&0\\
	 	\Psi'(\overline{w})&0&0&0&0
	\end{array}
	\right].
\end{equation}

Applying Schur's theorem and using various TC sum-rules listed
in appendix \ref{sec:atc}, one finds:  \begin{eqnarray}
	 Z^{(4)}\eq-\det\O^{(4)}\nonumber\\
	 \eq{1\over\pi^4}\det\left[
	 	\begin{array}{cccc}
			1&0&\e^{-R^2/2}&0\\
			0&1&R\e^{-R^2/2}&\e^{-R^2/2}\\
			\e^{-R^2/2}&R\e^{-R^2/2}&1&R\\
			0&\e^{-R^2/2}&R&1+R^2
		\end{array}\right]\nonumber\\
	\eq{1\over\pi^4}\left[\left(1-\e^{-R^2}\right)^2-R^4\e^{-R^2}\right].
	\label{i45}
\end{eqnarray}
To compare the correlation functions of divalent ions and simple
ions (in the bulk case),
we plot eq.(\ref{i26}) and (\ref{i45}) in Fig.\ \ref{fig:f2}.
Clearly, the divalent ions has a larger `exchange hole' near its
origin.

\section{Void Colloids in a Plasma}\label{chap:v} 
In this section, we introduce two empty colloidal voids into 
the plasma system.
We take the colloidal particle to be a circular disk with
finite radius $a$, inside which ions are excluded.
The  presence of the voids clearly destroys the system's
azimuthal symmetry. As a result, different angular momentum
orbitals intersect, and the overlap matrix is generally not diagonal.
It is the off-diagonal matrix elements that make the colloidal
interactions non-trivial.

We shall fix the two voids at $w$ and $\overline{w}$,
respectively.  Following eq.(\ref{pp5}), we can write down
the partition function of the system as the determinant of the
overlap matrix: 
\begin{equation}
	 Z=\det\O.\label{pp41}
\end{equation}
Because the regions occupied by colloids are not accessible to
the ions, their contribution to the overlap matrix should 
be excluded.  Thus, in contrast to eq.(\ref{pp9}), here we
have:\begin{equation}
	 O_{ij}=\langle i|j\rangle-\langle i|j\rangle_{w}-\langle
	 i|j\rangle_{\overline{w}}, \label{pp42}
\end{equation}
where $\langle i|j\rangle_w$ means a partial inner product integrated
over a circle with radius $a$ centered at $w$.
According to eq.(\ref{atc1}), eq.(\ref{pp42}) can be explicitly
evaluated as \cite{notes}:\begin{equation}
	 O_{ij}=\delta_{ij}-
	 \left[\sum_{l=1}^\infty\g{l}\Tb_{il}(w)\T_{jl}(w)+\cc\right],
	 \label{v1}
\end{equation}
where $\g{l}$ is the incomplete gamma function defined in
eq.(\ref{aoc3}).

\subsection{Analytic Results} \label{sec:va}
The off-diagonal matrix elements make the determinant in
eq.(\ref{pp41}) difficult to analyse.  However,
if the colloidal size $a$ is small,  we can treat them
perturbatively.  
From eq.(\ref{aoc3}) we know $\g{i}\sim
O(a^{2i})$ when $a\to 0$.
Thus, to the lowest order in $a$, we can truncate the 
infinite summation in eq.(\ref{v1}) at
$l=1$:\begin{equation}
	 O_{ij}\approx\delta_{ij}-\g{1}\left[\Tb_{i1}(w)\T_{j1}(w)
	 	+\T_{i1}(w)\Tb_{j1}(w)\right].\label{va1}
\end{equation}
If we define a $2\times n$ matrix $\J$ as \begin{equation}
	 \J\equiv\left[
	 	\begin{array}{cccc}
			\T_{11}(w)&\T_{21}(w)&\cdots&\T_{n1}(w)\\
			\Tb_{11}(w)&\Tb_{21}(w)&\cdots&\Tb_{n1}(w)
		\end{array}\right],
\end{equation}
eq.(\ref{va1}) may be rewritten in a matrix form:
\begin{equation}
	 \O=1-\J^\dagger\g{1}\J.
\end{equation}
According to the corollary of Schur's theorem (see appendix
\ref{chap:ast}, eq.(\ref{ast10})), we have:
\begin{equation}
	 \det\left(1-\J^\dagger\g{1}\J\right)_{n\times n}=
	 \det\left(1-\J\g{1}\J^\dagger\right)_{2\times 2}.
	 \label{va10}
\end{equation}
The determinant of a $2\times 2$ matrix is easy to calculate and
the result is (dropping terms with order higher than $\g{1}$):
\begin{equation}
	Z=\det\O\approx1-2\left(\sum_{k=1}^n|T_{k1}(w)|^2\right)\g{1}=1-2\g{1},
	\label{va20}
\end{equation}
where we have used the TC sum-rule (\ref{atc7}).
Notice that the above (first order) result is $R$-independent.
This is no accident, because in order for the colloids to  know 
the separation distance $R$,  the Green's function needs to be 
integrated over both colloidal regions, which yeilds a term proportional
to $a^4$, appearing in at least the second order in perturbation theory.

In the second order approximation, we must keep terms
proportional to $\g{1}$, $\g{1}^2$ and $\g{2}$.  
The procedure is quite similar to the above, and the result is:
\begin{eqnarray}
	Z&\approx&1-2\left(\sum_{k=1}^n|T_{k1}(w)|^2\right)\g{1}
	-2\left(\sum_{k=1}^n|T_{k2}(w)|^2\right)\g{2}\nonumber\\
	&&+\left[\left(\sum_{k=1}^n|T_{k1}(w)|^2\right)^2
		-\left|\sum_{k=1}^nT_{k1}^2(w)\right|^2\right]\g{1}^2.
		\label{pp49}
\end{eqnarray}
Compared to eq.(\ref{i15}) and (\ref{i25}),
we find that
the first and second order terms are respectively
proportional to the one- and two-body density functions.
Hence the colloidal void-void interaction does not
differ very much from the ion-ion interaction we studied earlier in 
Section \ref{chap:i}.

In the plasma bulk, we assume $w={\i{R\over2}}$ and use 
TC sum-rules (\ref{atc7}) and (\ref{atc8}) in eq.(\ref{pp49}) to obtain 
\begin{equation}
	 Z\approx1-2\g{1}-2\g{2}+\left(1-\e^{-R^2}\right)\g{1}^2;
\end{equation}
near the plasma edge(s), we assume $w=r\e^{\i\theta}$
($\theta={R\over2r}$) and express eq.(\ref{pp49}) in terms of TC 
sum-rules results:
\begin{equation}
	 Z^{s,d}\approx 1-2A^{s,d}_{11}\g{1}-2A^{s,d}_{22}\g{2}
	 	+\left[(A^{s,d}_{11})^2-|B^{s,d}_{11}|^2\right]\g{1}^2,
\end{equation}
where the explicit forms for $A^{s,d}$ and $B^{s,d}$ 
for asymptotically large $R$ can be found in eq.(\ref{atc11}), 
(\ref{atc14}), (\ref{atc21}) and (\ref{atc24}).

\subsection{Numerical Results} \label{sec:vn}
For large-sized colloids, the analytic results based on the small 
$a$-expansion may not be appropriate.
We resort to numerical methods and use LU factorization 
with partial pivoting to evaluate the matrix determinants.
We find that (essentially) exact results can be obtained from 
relatively small matrix sizes.

For colloidal voids immersed in the bulk of the plasma, our
numerical results are presented in Fig.\ \ref{fig:f3}.
For a variety of sizes we studied, no attraction is found,
confirming our analytic predictions.

Near the plasma edge(s), we find it more convenient to use 
square shaped colloids in numerical calculation.
We assume the  squares are of size $2a$, centered at 
$r\e^{\i\theta}$ and $r\e^{-\i\theta}$ respectively.
[The plasma edge(s) are located at $(r\pm{h\over2})$.]
Defining $\delta\equiv{a\over r}$, 
we can calculate the overlap matrix in eq.(\ref{pp42}) as 
following: \begin{equation}
	O_{ij}=\delta_{ij}-\left[
		\int_{r-a}^{r+a}\!\d\rho
		\int_{\theta-\delta}^{\theta+\delta}\!\d\phi\,
			{\rho^{i+j-1}\e^{-\rho^2}\e^{\i(j-i)\phi}
				\over\pi\sqrt{\Gamma(i)\Gamma(j)}}
		+\cc\right].
\end{equation}
In the limit $i\sim j\sim r^2\to\infty$,
the above integral can be evaluated asymptotically.  The result
is \begin{equation}
O_{ij}\approx\delta_{ij}-C(\theta,\delta)
{\Gamma({i+j\over 2})\over\sqrt{\Gamma(i)\Gamma(j)}}
	\left[\Phi\left(r_+\right)-\Phi\left(r_-\right)\right],
\end{equation}
where, $r_\pm\equiv{(r\pm a)^2-{i+j\over 2}+1\over\sqrt{i+j-2}}$;
$C(\theta,\delta)\equiv{\cos[(i-j)\theta]\sin[(i-j)\delta] 
\over(i-j)\pi}$ ($C(\theta,\delta)$ should be interpreted as 
${\delta\over\pi}$ in case of $i=j$).
In practice, we keep increasing $r$ until the numerical result 
becomes independent of $r$, which is found to occur approximately at
$r=60$. The results shown in Fig.\ \ref{fig:f4} (for single edged system) 
and Fig.\ \ref{fig:f5} (for double edged system) are computed using
$r=100$.  In both figures, when the colloidal voids are far 
away from the edge(s) ($h=6.0$), 
the curves are similar to those in Fig.\ \ref{fig:f3}, 
namely, the repulsive interaction has a Gaussian decay.  
As the colloids become closer to the plasma edge(s) (smaller $h$), 
the repulsion starts to develop a long tail, 
in agreement with our analytic results.

\section{Charged Colloids in a Plasma} \label{chap:c} 
We now consider charged colloids in the plasma.  
In the simplest case, the charged colloid is taken to be an
empty disk plus a fixed ion charge at the center. 
The ion-ion interaction and void-void interaction have been
studied in Section \ref{chap:i} and \ref{chap:v} respectively; 
both are repulsive.  However, as we shall see in this section, 
the ion-void interaction is usually attractive, and has a 
longer range.  The competition between the repulsion and 
attraction results in a richer behavior for the
charged colloidal interactions.

Following eq.(\ref{i20}) and (\ref{i21}), we have the 
partition function as follows:
\begin{eqnarray}
	 Z^{(2)}&=&-\det\O^{(2)}\\
	 \O^{(2)}&\equiv&\left[
	 \begin{array}{ccc}
	         \O&\Psi^\dagger(w)&
			 	\Psi^\dagger(\overline{w})\\
		     \Psi(w)&0&0\\
		     \Psi(\overline{w})&0&0
		   \end{array}
       \right],\label{pp51}
\end{eqnarray}
where, $\Psi\equiv(\psi_1,\psi_2,\ldots,\psi_n)$ is a row vector
representing the fixed ion charge at the colloidal center;
the $n\times n$ overlap matrix $\O$, according to eq.(\ref{v1}),
is non-diagonal:
\begin{equation}
	 O_{ij}=\delta_{ij}-
	 \left[\sum_{l=1}^\infty\g{l}\Tb_{il}(w)\T_{jl}(w)+\cc\right].
	 \label{c1}
\end{equation}

\subsection{Analytic Results} \label{sec:ca}
If the colloidal size $a$ is small, we can obtain an analytic
expression. 
One may truncate the infinite series in eq.(\ref{c1}) 
at $l=1$: \begin{equation}
	 O_{ij}\approx\delta_{ij}-\g{1}\left[\Tb_{i1}(w)\T_{j1}(w)
	 	+\T_{i1}(w)\Tb_{j1}(w)\right].
		\label{ca33}
\end{equation}
However we notice that the $(n+1)$th and $(n+2)$th row of matrix 
$\O^{(2)}$ in eq.(\ref{pp51}) can be written as \begin{eqnarray}
	 O^{(2)}_{(n+1)j}\eq\psi_j(w)={\T_{j1}(w)\over\sqrt{\pi}}\label{ca31}\\
	 O^{(2)}_{(n+2)j}\eq\ps_j(w)={\Tb_{j1}(w)\over\sqrt{\pi}}\label{ca32}.
\end{eqnarray}
If we multiply (\ref{ca31}) by
$\sqrt{\pi}\g{1}\Tb_{i1}(w)$, multiply (\ref{ca32}) by
$\sqrt{\pi}\g{1}\T_{i1}(w)$, and add them together to the $i$th
row, the off-diagonal part of $O_{ij}$ in eq.(\ref{ca33}) is cancelled
competely.  Thus the final answer does not contain $\g{1}$.
This is true even if we do not make any truncations in
eq.(\ref{ca33}).
Physically, it originates from the fact that the first angular 
momentum channel is occupied by the fixed ion,
so it is not accessible to the free ions.  
Another way of saying this is that the fixed ion serves as a
`Laughlin quasi hole' in the quantum electron system.

To obtain a non-trivial result, we consider the second 
angular momentum channel $l=2$: \begin{equation}
	 O_{ij}\approx\delta_{ij}-\g{2}\left[\Tb_{i2}(w)\T_{j2}(w)
	 	+\T_{i2}(w)\Tb_{j2}(w)\right].
		\label{ca0}
\end{equation}
Applying Schur's theorem,  we have:\begin{eqnarray}
	 Z^{(2)}\eq\det\O\cdot\det\left(
		  \left[\begin{array}{c}
		       \Psi(w)   \\\Psi(\overline{w})
		  \end{array}\right]\O^{-1}\left[
		  \begin{array}{cc}
		  	\Psi^{\dagger}(w)&\Psi^{\dagger}(\overline{w})
			\end{array}\right]\right)_{2\times 2}\nonumber\\
	&\equiv& Z_1\cdot Z_2.\label{ca3}
\end{eqnarray}
We first concentrate on the second factor $Z_2$.
It describes the two fixed ions (represented
by wavefunctions $\Psi$ and $\Psi^{\dagger}$) interacting through 
the plasma medium (represented by matrix $\O^{-1}$). 
For small $\g{2}$, the matrix $\O^{-1}$ can be approximated by:
\begin{equation}
	 O_{ij}^{-1}\approx\delta_{ij}+\g{2}\left[\Tb_{i2}(w)\T_{j2}(w)
	 	+\T_{i2}(w)\Tb_{j2}(w)\right],
\end{equation}
and $Z_2$ can be calculated to be (suppressing the $(w)$ arguments):
\begin{eqnarray}
	 Z_2&\approx&
	 \left[\left(\sum_{k=1}^n|\T_{k1}|^2\right)^2-
	 	\left|\sum_{k=1}^n\T_{k1}^2\right|^2\right]
	\nonumber\\
&&
	+2\g{2}\left[
		\left(\sum_{k=1}^n|\T_{k1}|^2\right)
		\left|\sum_{k=1}^n\T_{k1}\T_{k2}\right|^2
	\right.\nonumber\\
&&
		+\left(\sum_{k=1}^n|\T_{k1}|^2\right)
		\left|\sum_{k=1}^n\Tb_{k1}\T_{k2}\right|^2
	\nonumber\\
&& 
		-\left(\sum_{k=1}^n\Tb_{k1}^2\right)
		\left(\sum_{k=1}^n\T_{k1}\Tb_{k2}\right)
		\left(\sum_{k=1}^n\T_{k1}\T_{k2}\right)
\nonumber\\
&&\left. 
		-\left(\sum_{k=1}^n\T_{k1}^2\right)
		\left(\sum_{k=1}^n\Tb_{k1}\T_{k2}\right)
		\left(\sum_{k=1}^n\Tb_{k1}\Tb_{k2}\right)
	\right] \label{ca4}
\end{eqnarray}
We can interprete the above result by thinking
of the colloidal voids as a perturbation to the 
plasma system plus fixed ions. 
The first square bracket in eq.(\ref{ca4}) is the result for the 
unperturbed system, namely, the ion-ion interaction via the
plasma medium.  It is precisely
the two-body density function we obtained in eq.(\ref{i25}).
The second square bracket in eq.(\ref{ca4}) is the 
first order correction,
namely, a fixed ion on one site interacts with the colloidal void on
the {\it other} site.  This ion-void interaction is usually
found to be attractive, and since it involves higher order TC
sums, its range is longer than that of the ion-ion repulsion (see
results below).  So far, we have missed another first order effect, namely,
the fixed ion interacts with the colloidal void on the 
{\it same} site. 
The excess background charge underneath the void should
reduce the effective ionic charge somehow.
It turns out that this renormalization effect is 
captured precisely by 
the first factor $Z_1$ in eq.(\ref{ca3}): 
\begin{equation}
	 Z_1=\det\O\approx 1-2\sum_{k=1}^{n}|\T_{k2}|^2\g{2}.
\end{equation}

In the bulk of plasma, we assume $w=\i{R\over2}$ and use 
TC sum-rules (\ref{atc7}) and (\ref{atc8}) to obtain:
\begin{equation}
	Z^{(2)}\approx(1-2\g{2})\left(1-\e^{-R^2}\right)
		+2R^2\e^{-R^2}\g{2}.\label{ca40}
\end{equation}
In the above equation, the first term is the renormalized 
ion-ion repulsion and the second term is the ion-void
attraction.  The attraction is proportional to $R^2\e^{-R^2}$,
which has a longer range than the repulsion.  This feature is not
found in mean field approximations,  thus must be caused by the
strong correlation effects.
Notice that the void-void repulsion and the renormalization 
of the ion-void attraction are both of second order, so they do 
not appear in eq.(\ref{ca40}).
The repulsion and attraction respectively dominates the short
distance and long distance behavior,
and the minimum free energy $\beta F=-\ln Z^{(2)}$ occurs
where the first derivative vanishes:
\begin{equation}
	{\d Z^{(2)}(R^*)\over\d R}=0\quad{\mathrm or}\quad 
	R^*=\sqrt{1\over 2\g{2}}.\label{ca6}
\end{equation}
Fig.\ \ref{fig:f6} plots the calculated free energy vs. $R$ for various
values of $a$.
Note that when $\g{2}$ is small ($a=0.6$),
eq.(\ref{ca6}) predicts that the location
of the free energy minimum is far away from origin, in which case
the attraction is strongly suppressed by the decaying factor
$\e^{-R^2}$, thus becoming virtually invisible.  

Near the plasma edge, we 
use TC sum-rules (\ref{atc11}) through (\ref{atc15}) to obtain
(for asymptotically large $R$):
\begin{eqnarray}
 	Z^{(2)s}&\approx&\left[(A_{11}^s)^2-{\e^{-h^2}\over2\pi R^2}\right]
		\nonumber\\
&&+
	 	2\g{2}\left[{(A_{11}^sh^2+A_{22}^s)\e^{-h^2}
			-{2h\over\sqrt{2\pi}}\e^{-{3h^2\over 2}}
		\over2\pi R^2}\right.\nonumber\\
&&+
		\left.A_{11}^s\left({\e^{-h^2}\over\sqrt{2\pi}}-A_{11}^sA_{22}^s\right)
		\right],
\end{eqnarray}
Similarly for double edged system, we use TC sum-rules (\ref{atc21})
through (\ref{atc25}) to obtain
\begin{eqnarray}
	 Z^{(2)d}&\approx&\left[(A_{11}^d)^2-{\e^{-h^2}\over2\pi R^2}\sin^2(hR)\right]
\nonumber\\
&&+
	 	2\g{2}\left\{{[A_{11}^dh^2\cos^2(hR)+A_{22}^d\sin^2(hR)]
		\e^{-h^2}\over 2\pi R^2}\right.\nonumber\\
&&-
		\left.(A_{11}^d)^2A_{22}^d \right\},\quad\quad\quad
		\label{osc}
\end{eqnarray}

\subsection{Numerical Results} \label{sec:cn}
Our numerical results for charged colloids in the bulk case
are plotted in Fig.\ \ref{fig:f7}.
When $a$ is small ($a=0.6-0.8$)  the numerical results 
agree very well with the analytic ones presented in Fig.\ \ref{fig:f6}.
However, when $a$ is large ($a=1.0$), the two differ
significantly in that the
numerical free energy has an additional secondary maximum.
This peak reflects the even longer ranged, second order 
effects which are not accounted for in the analytic
studies.

We also tried putting double ion charges at the center of 
each colloids.  This case is very similar to the divalent ions which 
are studied in Section \ref{sec:i4}.
Our numerical results are presented in Fig.\ \ref{fig:f8},  where
the colloidal sizes are chosen to be $\sqrt{2}$ times as
large as those in Fig.\ \ref{fig:f7}, so that the two figures
may be comparable.
We see that the qualitative behavior is very similar.
The attraction minimum stays further out.

Near the plasma edge, we use 
squares to represent colloidal particles as we did in Section
\ref{sec:vn}.
The numerical results for two different sizes of colloids are
presented in Fig.\ \ref{fig:f9} and Fig.\ \ref{fig:f10}.
In case I ($a=0.8$, Fig.\ \ref{fig:f9}), when the colloids are 
far away from the edge ($h=5.0$) they repel; as they become
closer to the edge ($h=2.3$) they start to attract each other 
within a certain range ($R<2.25$), beyond that range they still
repel; at a particular point ($h=1.5$), the range of the
attraction seems to extend to infinity;  finally, when they are
too close to the edge ($h=0.8$), due to insufficient screening, 
a long range repulsion is found.
In case II ($a=0.9$, Fig.\ \ref{fig:f10}), the colloids are 
larger in size and  attract each other even in the bulk ($h=5.0$);
the attraction becomes deeper ($h=2.0$)
and longer ranged ($h=1.3$) when the colloids move towards the
edge. 

Fig.\ \ref{fig:f11} shows the free energy of colloids in a 
double edged system.  When the two boundaries are widely separated
($h=5.0$), the colloids repel;  as the separation decreases,
the free energy develops a long oscillating tail.  These are
consistent with the analytic results.
Note that the oscillation `wavelength', according to eq.(\ref{osc}),
is set by the strip width $h$, and is independent of the size $a$ 
of the colloids.  
This conclusion is confirmed by Fig.\ \ref{fig:f12}.

Finally, Fig.\ \ref{fig:f13} shows the results for doubly
charged colloids near a single plasma edge.  They are qualitatively 
similar to those shown in Fig.\ \ref{fig:f10}.

\subsection{Where Does Colloid Stay?} \label{sec:c1}
To better understand the charged colloidal interactions,
especially near the plasma edge,
it is instructive to learn where  they
want to stay in the plasma.
We put a charged colloid at $w=r\e^{\i\theta}$, which is
${h\over2}$ away from the plasma edge at $(r+{h\over2})$, 
and  calculate the
partition function to be: \begin{eqnarray}
	 Z^{(1)s}&=&-\det\O^{(1)}\\
	 \O^{(1)}&=&
	 	\left[\begin{array}{cc}
			\O&\Psi^\dagger(w)\\
			\Psi(w)&0
			\end{array} \right],
\end{eqnarray}
where the $n\times n$ overlap matrix $\O$ is:
\begin{equation}
	 O_{ij}=\delta_{ij}-\sum_{l=1}^\infty\g{l}\Tb_{il}(w)\T_{jl}(w).
	 \label{c17}
\end{equation}
It can be shown that the first angular momentum channel is 
blocked by the fixed
ion charge so that the final answer does not depend on $\g{1}$.
For small $a$, we truncate the infinite summation in eq.(\ref{c17}) 
at $l=2$ and use Schur's theorem to find:
\begin{eqnarray}
	Z^{(1)s}&\approx&\left[1-\sum_{i=1}^n|\T_{i2}|^2\g{2}\right]
	   \left[\sum_{i=1}^n|\T_{i1}|^2+
	   \left|\sum_{i=1}^n\Tb_{i1}\T_{i2}\right|^2\g{2}
					   \right]\nonumber\\
	\eq A_{11}^s+\left({\e^{-h^2}\over2\pi}-A_{11}^sA_{22}^s\right)\g{2}.
\end{eqnarray}

Fig.\ \ref{fig:f14} shows the numerical results done for a
square charged colloid.  We plot the free energy $\beta F$ 
as a function of $h$ for 
three different sized squares.  
In all cases,
we find the colloid is attracted to a free energy minimum near 
the plasma edge (at $h=0$).
For $h<0$, the free energy quickly blows up.  This tells that
the colloid does not want to flee away from the plasma
edge,  because its fixed charge is attracted by
the background charge and wants to be in the
lowest available angular momentum channel.
The optimal distance $h^*$ is in the range in which attraction
between colloids occurs.

\section{Summary}
In this paper, we considered a strongly coupled ($\beta=2$) 2D 
one-component plasma model and calculated the effective 
interaction between two colloids both in the bulk and near
boundaries.
We found that due to ion-void interaction, the colloids may attract
despite the fact that ion-ion and void-void interactions are
both repulsive.  Near the plasma edge, the attraction effects
become more prominent.  Though our model is oversimplified, the
exact results provide useful insights into the problem, and can
serve as a testing ground for more sophisticated approximate
treatments.

\section*{Acknowledgement}
We thank Adrian Parsegian and Rudi Podgornik for introducing us
to this problem.  RR is grateful to SMG and the Physics Department
of Indiana University for supporting a summer visit during which
this work was initiated. SMG and RR are grateful to the 
Institute for Theoretical Physics at UCSB where part
of this work was performed.  This work is supported by 
NSF DMR-9714055.
	
\appendix
\section{Properties of Lowest Landau Level Wavefunctions}\label{chap:alw}
Some important properties of lowest Landau level (LLL) wavefunctions 
are summarized here.   
They are frequently used throughout the paper.

\subsection{Orthonormality and Completeness} \label{sec:aoc}
The LLL wavefunctions defined in eq.(\ref{pm3}) are 
orthonormal:
\begin{equation}
	 \langle i|j\rangle = \int\!\d
	 z\,\ps_i(z)\psi_j(z)=\delta_{ij}.\label{aoc1}
\end{equation}
Sometimes, we want to calculate a partial inner
product (the overlap integrated over part of the entire plane),
and the result is generally not diagonal.  
However, if the integration region
has azimuthal symmetry, the conservation of angular momentum
guarantees an exception.  As an example, let us take the region to be
a circle with radius $a$ centered at origin (denoted by the
subscript),
\begin{equation}
	 \langle i|j\rangle_0 = \int_0^a\!\rho\d\rho
	 \int_0^{2\pi}\!\d\phi
	 {\rho^{i+j-2}\e^{-\rho^2}\e^{\i(j-i)\phi}\over
	     \pi\sqrt{\Gamma(i)\Gamma(j)}} =\delta_{ij}\g{i},
		 \label{aoc2}
\end{equation}
where the imcomplete gamma function $\g{i}$ is defined as the following:
\begin{equation}
     \g{i}\equiv{1\over\Gamma(i)}\int_0^{a^2}\!\d u\,\e^{-u}u^{i-1}
	     =1-\e^{-a^2}\left(\sum_{k=0}^{i-1}{a^{2k}\over k!}\right).
		 \label{aoc3}
\end{equation} 
Notice that in the limit $a\to 0$, $\g{i}\sim O(a^{2i})$.

The completeness tells that any state lying in LLL can be 
expressed as a linear combination
of these basis wavefunctions.  An example is
the wavefunction $\psi_i(z)$ with its argument translated by a 
vector $w$: \begin{equation}
   \e^{{1\over 2}(\overline{z}w-z\overline{w})}\psi_i(z+w)
       =\sum_{j=1}^\infty \T_{ij}(w)\psi_j(z). \label{aoc4}
\end{equation}
Here, the extra pure phase factor in front of $\psi_i(z+w)$ 
is needed to project the translated wavefunction into the LLL.

\subsection{Translation Coefficients}
\label{sec:atc}
The linear coefficients $\T_{ij}$s defined in eq.(\ref{aoc4}) 
are termed {\it translation coefficients} (TCs). 
Setting $z=0$ in eq.(\ref{aoc4}), we find:
\begin{equation}
	 \psi_i(w)=\sum_{j=1}^\infty\T_{ij}(w)\psi_j(0)={\T_{i1}(w)\over\sqrt{\pi}}.
	 \label{atc0}
\end{equation}
Thus the LLL basis functions are just special cases of TCs.
Another useful application of TC is to calculate the partial
inner product on a circle with radius $a$ centered at $w\neq 0$ 
(denoted by the subscript): 
\begin{eqnarray}
	\langle i|j\rangle_w\eq\int_{|z|<a}\d z\,
	 	\ps_i(z+w)\psi_j(z+w)\nonumber\\
	\eq\int_{|z|<a}\d z
		\sum_{k,l}\Tb_{ik}(w)
		\ps_k(z)\psi_l(z) \T_{jl}(w)\nonumber\\
	\eq\sum_{k,l}\Tb_{ik}(w)\langle
		k|l\rangle_0\T_{jl}(w)\nonumber\\
	\eq\sum_{l=1}^\infty\g{l}\Tb_{il}(w)\T_{jl}(w).
	\label{atc1}
\end{eqnarray}

The TCs can be evaluated explicitly.
We multiply $\ps_k(z)$ to both sides of eq.(\ref{aoc4}),
and integrate $z$ over the entire plane. 
On the right hand side, this
picks out one particular $\T_{ik}$ from the summation
because of the orthonormality:
\begin{eqnarray}
	\T_{ik}(w)\eq\int\!\d z\,\e^{{1\over 2}(\overline{z}w-z\overline{w})}
		\overline{\psi}_k(z)\psi_i(z+w) \nonumber\\
	\eq{\e^{-|w|^2/2}\over\pi\sqrt{\Gamma(i)\Gamma(k)}} \int\!\d z\,
		\e^{-|z|^2-z\overline{w}}\overline{z}^{k-1}(z+w)^{i-1}\nonumber\\
	\eq{\e^{-|w|^2/2}\over\pi\sqrt{\Gamma(i)\Gamma(k)}} \int\!\d z\,
		\e^{-|z|^2}\nonumber\\
	&&\times\left[\sum_{t=0}^\infty{(-\overline{w})^tz^t\over t!}\right]
		\overline{z}^{k-1}
		\left[\sum_{s=1}^iC^{i-1}_{s-1}z^{s-1}w^{i-s}\right]\nonumber\\
	\eq\sqrt{\Gamma(i)\Gamma(k)}(-1)^k\hat{w}^{i-k}\e^{-{|w|^2\over 2}}
		\nonumber\\
	&&\times\sum_{s=1}^{\min(i,k)}{(-1)^s|w|^{i+k-2s}\over
		\Gamma(s)(i-s)!(k-s)!}\label{atc2}. 
\end{eqnarray}   
Where, in the 3rd equation we used Taylor and binomial expansions,
and in the 4th equation we used orthonormality (\ref{aoc1}).
Eq.(\ref{atc2}) immediately tells that the TCs have the following inversion
symmetry: \begin{equation}
	 \T_{ij}(w)=\Tb_{ji}(-w)=\T_{ji}(-\overline{w}).
	 \label{atc3}
\end{equation}

Another important property of TC arises when one composes two
successive translations: \begin{equation}
	 \T_{ij}(w_1+w_2)=\e^{{1\over 2}(\overline{w}_1w_2-w_1\overline{w}_2)}
	 	\sum_{k=1}^\infty\T_{ik}(w_1)\T_{kj}(w_2).
		\label{atc4}
\end{equation}
Or using eq.(\ref{atc3}), this becomes:\begin{eqnarray}
	 \T_{ij}(w_1+w_2)
	 	\eq\e^{{1\over 2}(\overline{w}_1w_2-w_1\overline{w}_2)}
	 		\sum_{k=1}^\infty\Tb_{ki}(-w_1)\T_{kj}(w_2)\label{atc5}\\
	 	\eq\e^{{1\over 2}(\overline{w}_1w_2-w_1\overline{w}_2)}
	 		\sum_{k=1}^\infty\T_{ki}(-\overline{w}_1)\T_{kj}(w_2)
			\label{atc6}
\end{eqnarray}
As a simple application, we set $-w_1=w_2=w$ in
eq.(\ref{atc5}): \begin{equation}
	 \lim_{n\to\infty}\sum_{k=1}^n\Tb_{ki}(w)\T_{kj}(w)
	 	=\T_{ij}(0)=\delta_{ij}, \label{atc7}
\end{equation}
or set $w_1=w_2={{\i R\over 2}}$ in eq.(\ref{atc6}): 
\begin{equation}
	 \lim_{n\to\infty}\sum_{k=1}^n\T_{ki}\left({\i R\over 2}\right)
	 \T_{kj}\left({\i R\over 2}\right)=\T_{ij}(\i R).\label{atc8}
\end{equation}
Note that if we interprete $\T_{ij}$ as the matrix elements of
$\Tm$, then eq.(\ref{atc7}) has a simple matrix form:
\begin{equation}
	 \Tm^\dagger(w)\Tm(w)=1, \label{atcr}
\end{equation}
showing that matrix $\Tm$ is unitary.  In the same matrix
notation, eq.(\ref{atc3}) is rewritten as\begin{equation}
	 \Tm^\dagger(w)=\Tm(-w)
\end{equation}
which along with eq.(\ref{atcr}) makes sense since $\Tm(-w)$
should be the inverse of $\Tm(w)$.

Eq.(\ref{atc7}) and (\ref{atc8}) are examples of TC
sum-rules.  They are frequently used in analysing the bulk of
the plasma.  Near the edge, we often encounter TC summations
of a different type:
\begin{eqnarray}
	 A_{ij}^{s}&\equiv&\lim_{r\to\infty}\sum_{k=1}^{(r+{h\over2})^2}
	 	\Tb_{ki}(r\e^{\i\theta})
	 	\T_{kj}(r\e^{\i\theta})\\
	 B_{ij}^{s}&\equiv&\lim_{r\to\infty}
	 	\sum_{k=1}^{(r+{h\over2})^2}
	 	\T_{ki}(r\e^{\i\theta})
	 	\T_{kj}(r\e^{\i\theta}).
\end{eqnarray}
Here, the superscript `s' denotes
single edged system and $\theta\equiv {R\over2r}$.
The difference between the bulk TC sum and edge TC sum is as
following:
treated as a perturbation, colloids in the
bulk mainly perturb the low angular momentum channels.
Such a perturbation has a discrete characteristic,
namely only a small number of channels are strongly affected.
Colloids near the boundary, on the other hand, perturb the
high angular momentum channels.  The perturbation has a
continuum nature, namely there are a large number of channels 
($r^2-hr<k<r^2+hr$) being weakly affected.
As a result, the evaluation of edge TC sum inevitably requires
one to convert the discrete summation into a continuum integral.
For example, let us calculate $A_{11}^s$: \begin{eqnarray}
	 A_{11}^{s}\eq\lim_{r\to\infty}\sum_{k=1}^{(r+{h\over2})^2}
		{r^{2k-2}\over\Gamma(k)}\e^{-r^2}\nonumber\\
	&\approx&\lim_{r\to\infty}\int_0^{r^2+hr}\!{\d x\over\sqrt{2\pi x}}\,
		\e^{2x\ln r+x-x\ln x-r^2},\label{atc10}
\end{eqnarray}
where we have defined $x\equiv k-1$ and used Stirling's asymptotic
formula: \begin{equation}
	 \Gamma(x+1)\approx\sqrt{2\pi x}\,\e^{-x+x\ln x}.
\end{equation}
The integal in eq.(\ref{atc10}) can be evaluated using the saddle 
point approximation, in which one expands the exponent near 
its minimum at $x=r^2$ up to the second order: \begin{equation}
	 A_{11}^{s}\approx\lim_{r\to\infty}\int_0^{r^2+hr}\!
	 	{\d x\over\sqrt{2\pi}r}\e^{-{1\over2r^2}(x-r^2)^2},
\end{equation}
substituting $y\equiv{x-r^2\over r}$, one finds\begin{equation}
	 A_{11}^{s}=\int_{-\infty}^h\!{\d y\over\sqrt{2\pi}}\e^{-y^2/2}
	 	={1\over 2}\left[1+\Phi\left({h\over\sqrt{2}}\right)\right].
		\label{atc11}
\end{equation}
Where $\Phi(x)$ is the standard error function.  
Other TC sum rules can be obtained through quite similar
procedures.  We omit the tedious derivations and 
simply list the results below: \begin{eqnarray}
A_{22}^{s}\eq{1\over 2}\left[1+\Phi\left({h\over\sqrt{2}}\right)
			-{h\e^{-h^2/2}\over\sqrt{2\pi}}\right], \label{atc12}\\
A_{12}^{s}\eq{\e^{-h^2/2}\over\sqrt{2\pi}}, \label{atc13}\\
B_{11}^{s}\eq{\e^{-h^2/2}\over\sqrt{2\pi}R}
			\e^{\i(hR-{\pi\over2})},\label{atc14}\\
B_{12}^{s}\eq{h\e^{-h^2/2}\over\sqrt{2\pi}R} 
			\e^{\i(hR-{\pi\over2})}.\label{atc15}
\end{eqnarray} 
Note that the above results are correct only
asymptotically for $R\to\infty$ ($R\equiv 2r\theta$).

In a double edged system, we encounter the same type of TC
summation, except that the lower limit  is changed from $k=1$ to
$k={(r-{h\over2})^2}$:  
\begin{eqnarray}
	 A_{ij}^{d}&\equiv&\lim_{r\to\infty}
	 	\sum_{k=(r-{h\over2})^2}^{(r+{h\over2})^2}
	 	\Tb_{ki}(r\e^{\i\theta})
	 	\T_{kj}(r\e^{\i\theta})\\
	 B_{ij}^{d}&\equiv&\lim_{r\to\infty}
	 	\sum_{k=(r-{h\over2})^2}^{(r+{h\over2})^2}
	 	\T_{ki}(r\e^{\i\theta})
	 	\T_{kj}(r\e^{\i\theta}).
\end{eqnarray}
Correspondingly, we have: (the
superscript `d' denotes double edges):
\begin{eqnarray}
A_{11}^{d}\eq\Phi\left({h\over\sqrt{2}}\right), \label{atc21}\\
A_{22}^{d}\eq\Phi\left({h\over\sqrt{2}}\right)
		-{h\e^{-h^2/2}\over\sqrt{2\pi}}, \label{atc22}\\
A_{12}^{d}\eq 0, \label{atc23}\\
B_{11}^{d}\eq {\e^{-h^2/2}\over\sqrt{2\pi}R}\sin(hR)
			\e^{\i(hR-{\pi\over2})},\label{atc24}\\
B_{12}^{d}\eq {h\e^{-h^2/2}\over\sqrt{2\pi}R}\cos(hR)
			\e^{\i(hR-{\pi\over2})}.\label{atc25}
\end{eqnarray}

\section{Schur's Theorem}\label{chap:ast}
For any four matrices $\A_{r\times r}$, $\B_{r\times s}$,
$\C_{s\times r}$ and $\D_{s\times s}$,  if  $\A$
is not singular,  Schur's theorem states\begin{equation}
	 \det\left[\begin{array}{cc}
	 	\A&\B\\\C&\D\end{array}\right]=\det
		\A\cdot\det(\D-\C\A^{-1}\B).
		\label{ast1}
\end{equation}
The proof is straightforward.  We consider the following matrix
identity: \begin{equation}
	 \left[\begin{array}{cc}1_{r\times r}&0_{r\times s}\\
	 	-\C\A^{-1}&1_{ s\times s}\end{array}\right]
	 \left[\begin{array}{cc}\A&\B\\
	 	\C&\D\end{array}\right]=
	 \left[\begin{array}{cc}\A&\B\\
	 	0_{ s\times r}&\D-\C\A^{-1}\B\end{array}\right].
		\label{ast2}
\end{equation}
Taking the determinant of both sides of eq.(\ref{ast2}),
eq.(\ref{ast1}) follows immediately.

Schur's theorem has a very useful corollary:
for any matrices $\J_{r\times s}$ and $\K_{s\times r}$, one has
\begin{equation}
	 \det(1-\J\K)_{r\times r}=\det(1-\K\J)_{s\times s}. 
	 \label{ast10}
\end{equation}
We can  prove this via very similar method.  First we consider
\begin{equation}
	 \left[\begin{array}{cc}1_{r\times r}&-\J\\
	 	0_{s\times r}&1_{ s\times s}\end{array}\right]
	 \left[\begin{array}{cc}1_{r\times r}&\J\\
	 	\K&1_{s\times s}\end{array}\right]=
	 \left[\begin{array}{cc}(1-\J\K)_{r\times r}&0_{r\times s}\\
	 	\K&1_{s\times s}\end{array}\right],
\end{equation}
which tells \begin{equation}
	 \det\left[\begin{array}{cc}1_{r\times r}&\J\\
	 	\K&1_{s\times
		s}\end{array}\right]=\det(1-\J\K)_{r\times r};
		\label{ast7}
\end{equation}
on the other hand, if we consider
\begin{equation}
	 \left[\begin{array}{cc}1_{r\times r}&0_{r\times s}\\
	 	-\K&1_{ s\times s}\end{array}\right]
	 \left[\begin{array}{cc}1_{r\times r}&\J\\
	 	\K&1_{s\times s}\end{array}\right]=
	 \left[\begin{array}{cc}1_{r\times r}&\J\\
	 	0_{ s\times r}&1-\K\J\end{array}\right],
\end{equation}
we obtain \begin{equation}
	 \det\left[\begin{array}{cc}1_{r\times r}&\J\\
	 	\K&1_{s\times
		s}\end{array}\right]=\det(1-\K\J)_{s\times s}.
		\label{ast8}
\end{equation}
A combination of eq.(\ref{ast7}) and (\ref{ast8}) results in
 eq.(\ref{ast10}).
 Letting $\J=\g{1}\K^{\dagger}$ in eq.(\ref{ast10}), we
obtain eq.(\ref{va10}).


\myfigure{6cm}{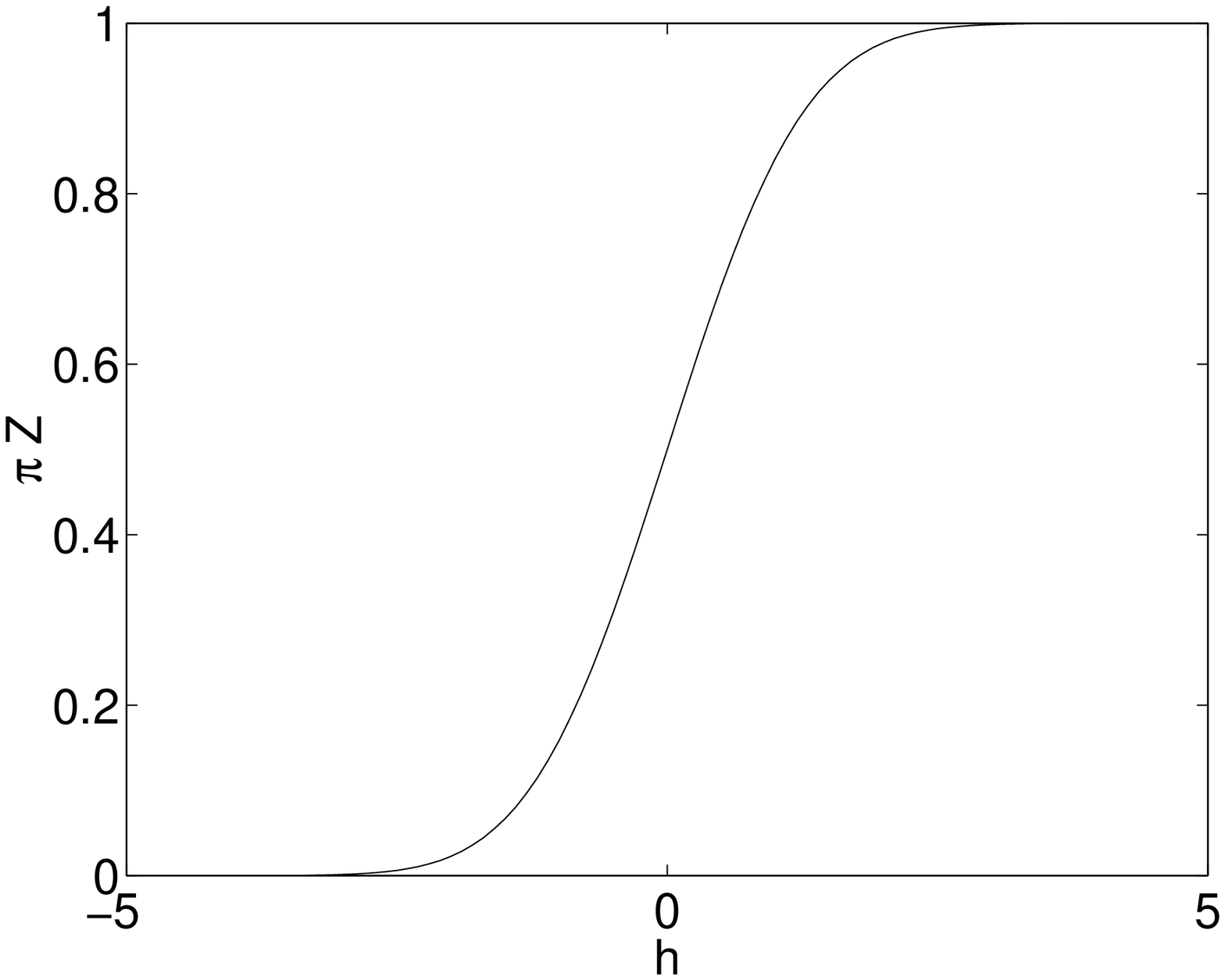}{The one-body density function near the plasma edge.}
\myfigure{6cm}{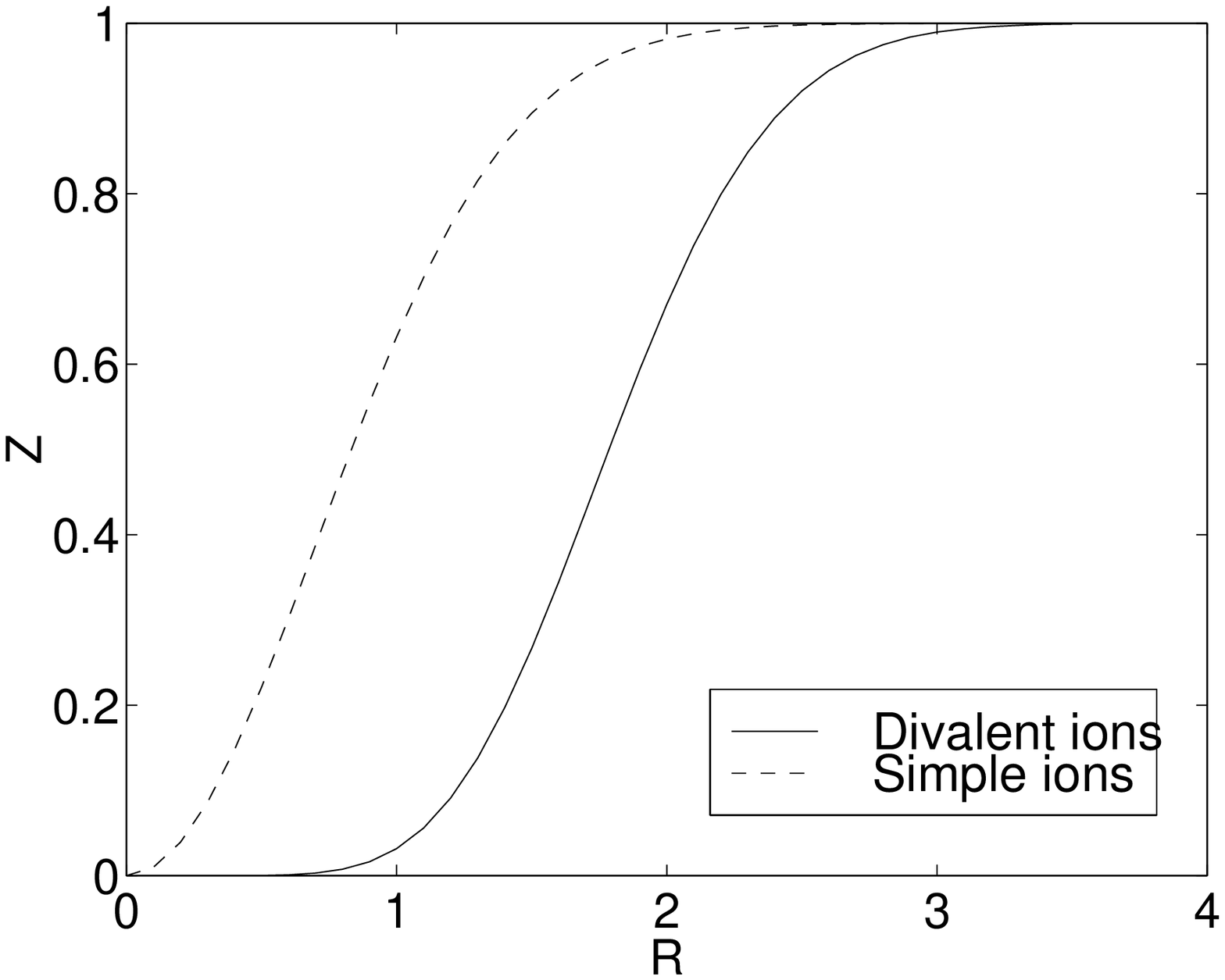}{The comparison between the two-body
density functions for divalent ions (the solid line) and
simple ions (the dash line) in the bulk.}
\myfigure{6cm}{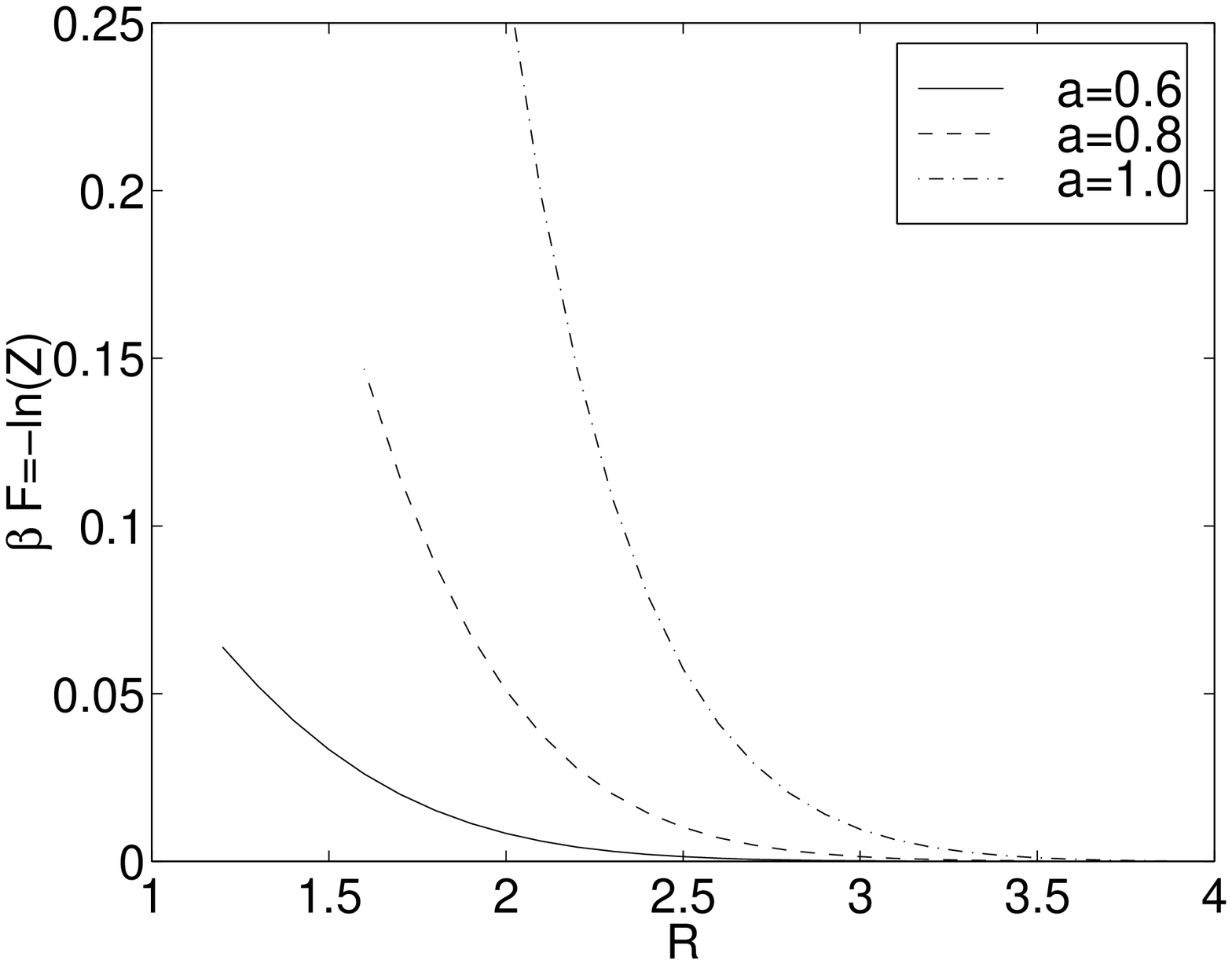}{Numerical results for the free energy of
colloidal voids of radius $a$ in the bulk of plasma separated by
a distance $R$.}
\myfigure{6cm}{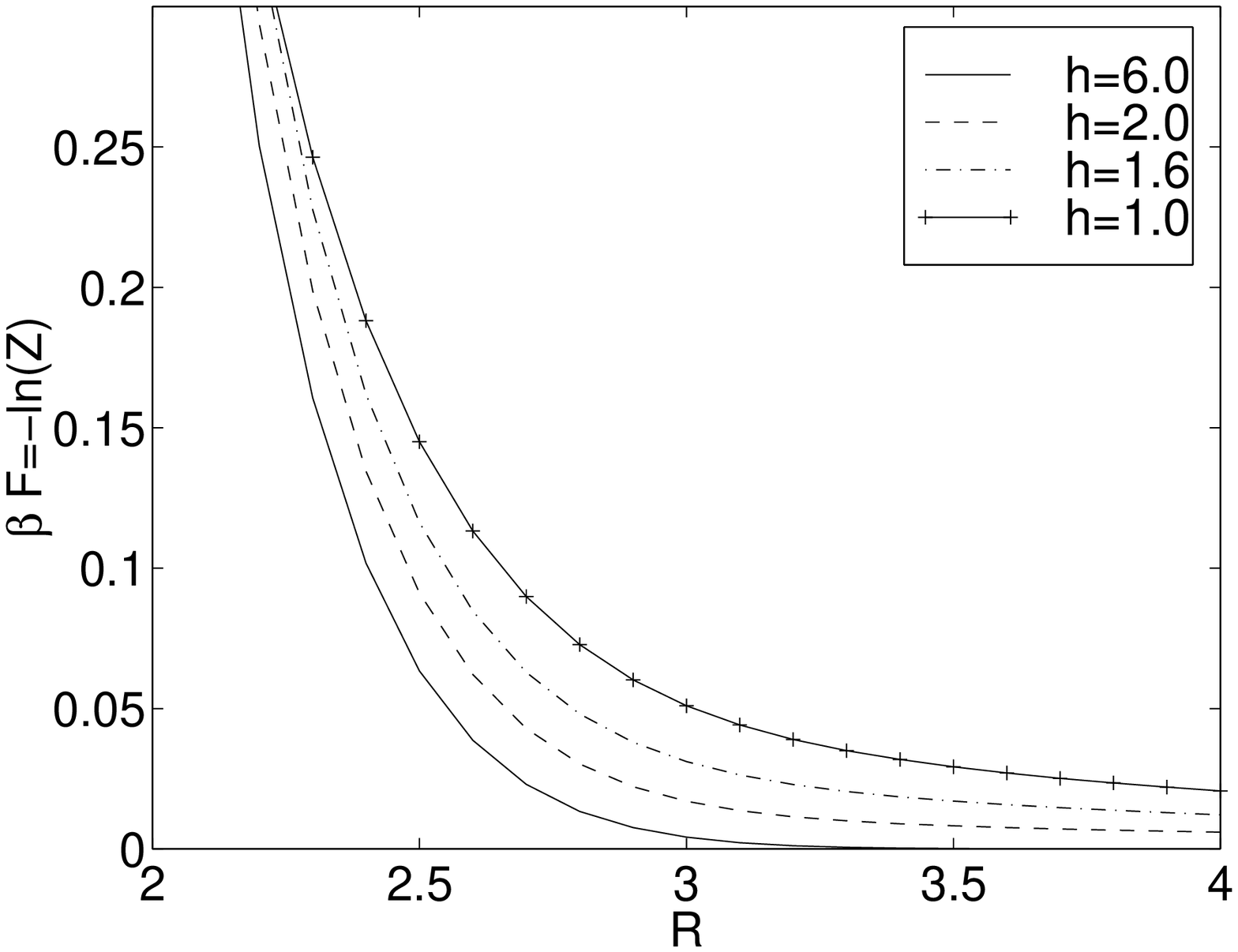}{Numerical results for the free energy of
empty colloids of radius $a=1.0$ and located a distance $h/2$ 
away from a single plasma edge as a function of their 
separation $R$.}
\myfigure{6cm}{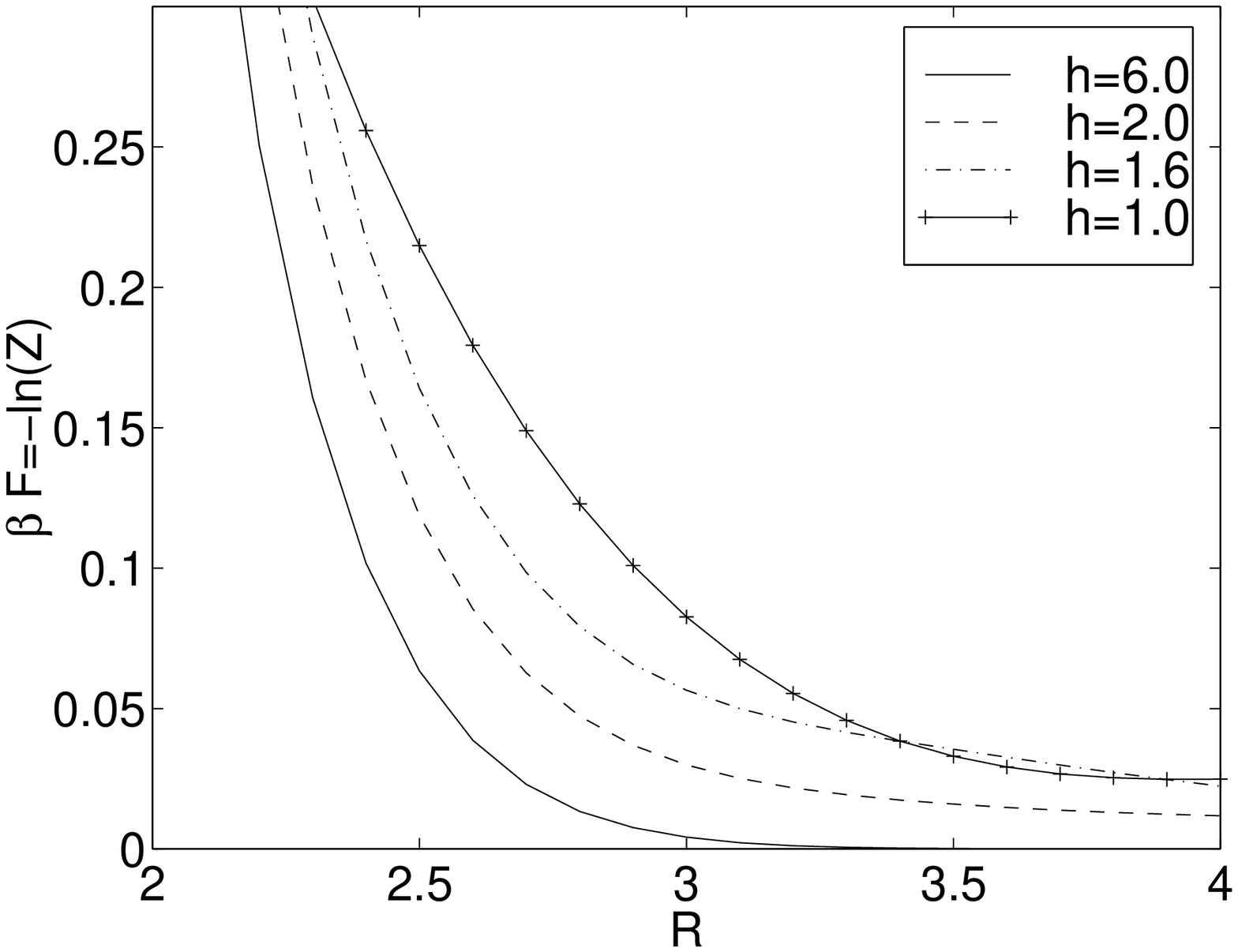}{Numerical results for the free energy of
empty colloids of radius $a=1.0$ separated by a distance $R$ in 
a plasma strip of width $h$.}
\myfigure{6cm}{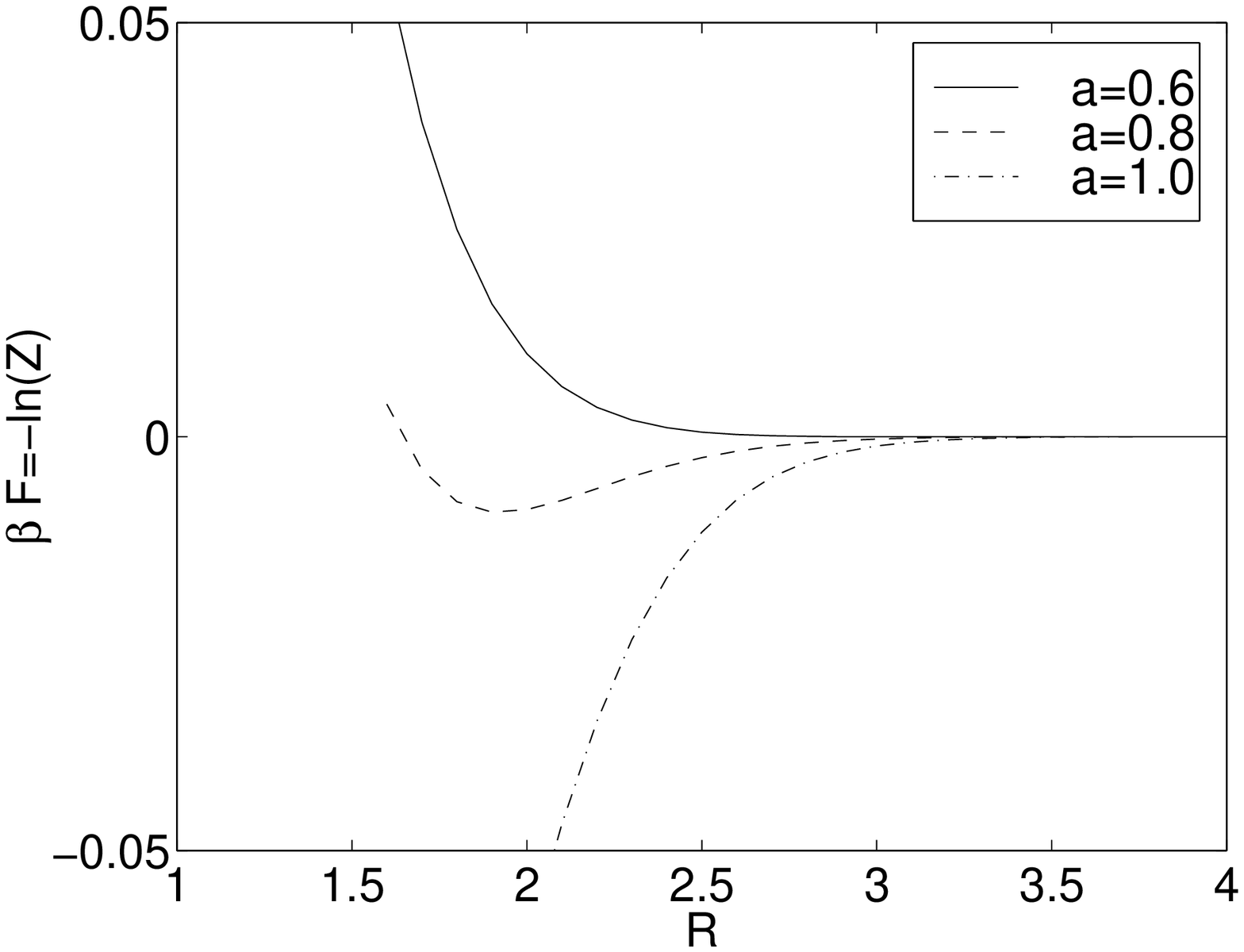}{Analytic results for the free energy vs. 
separation distance $R$
of singly charged colloids in the bulk of the plasma.}
\myfigure{6cm}{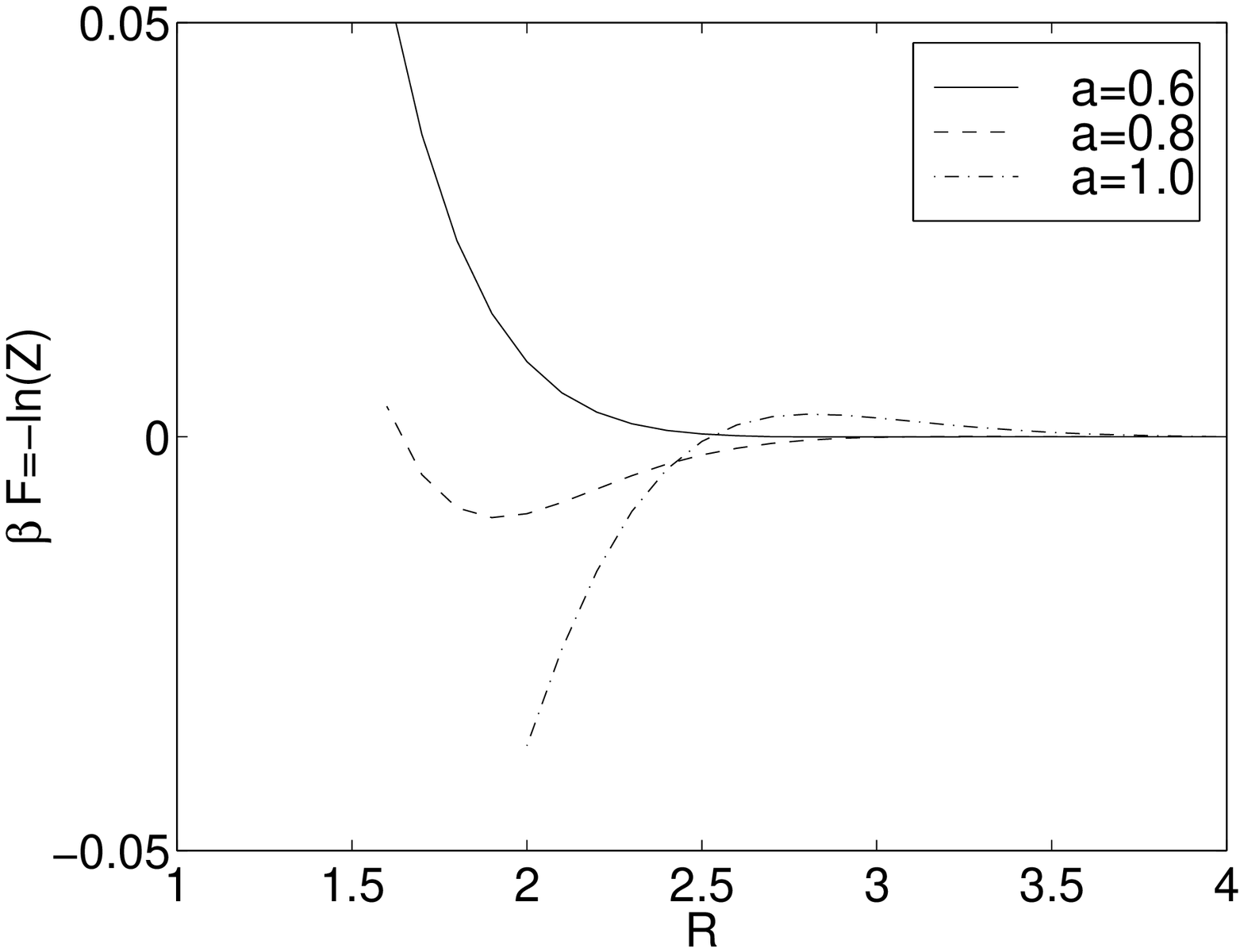}{Numerical results for the free energy vs.
separation distance $R$
of singly charged colloids of radius $a$ in the bulk of the plasma.}
\myfigure{6cm}{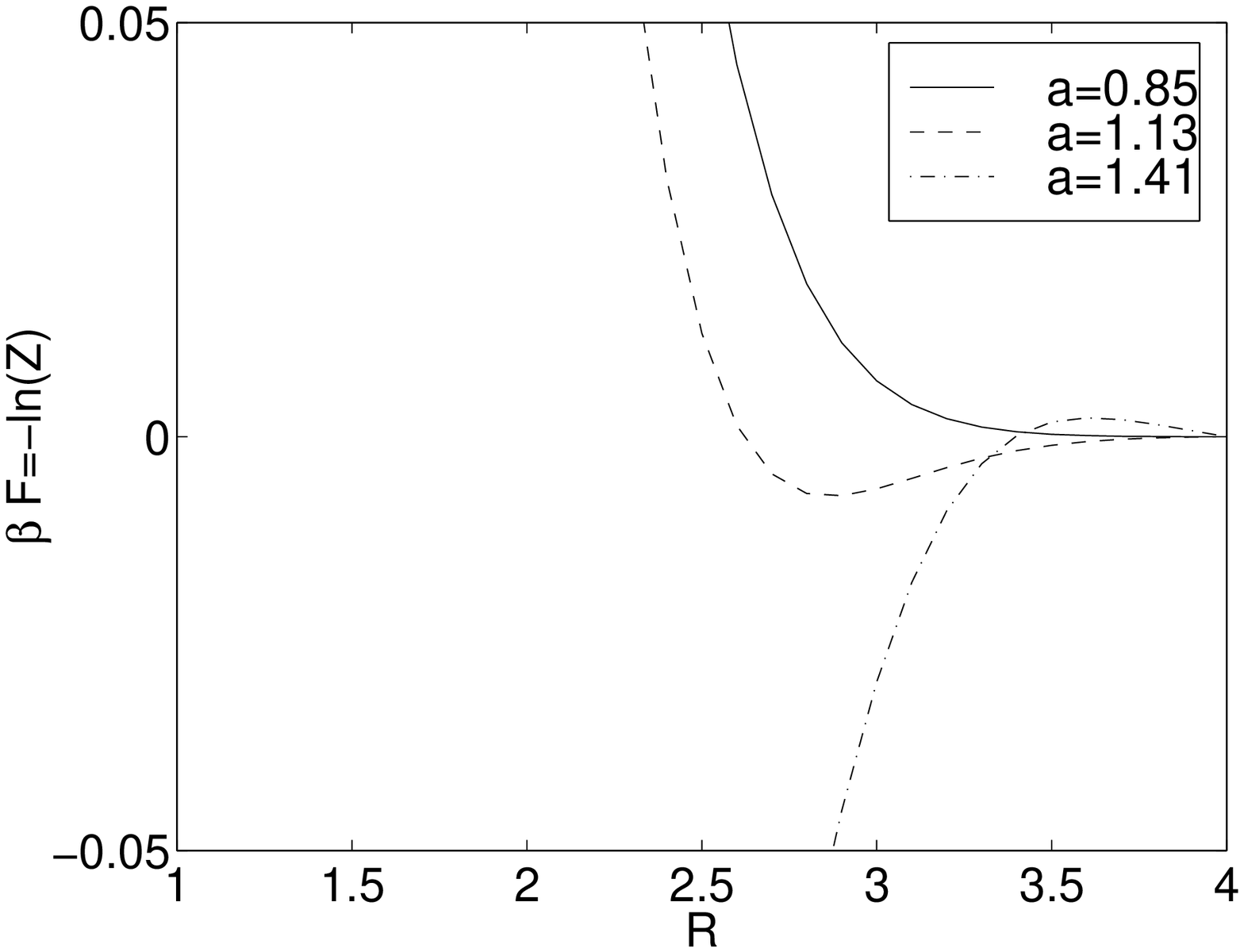}{Numerical results for the free energy vs.
separation distance $R$
of doubly charged colloids of radius $a$ in the bulk of the plasma.}
\myfigure{6cm}{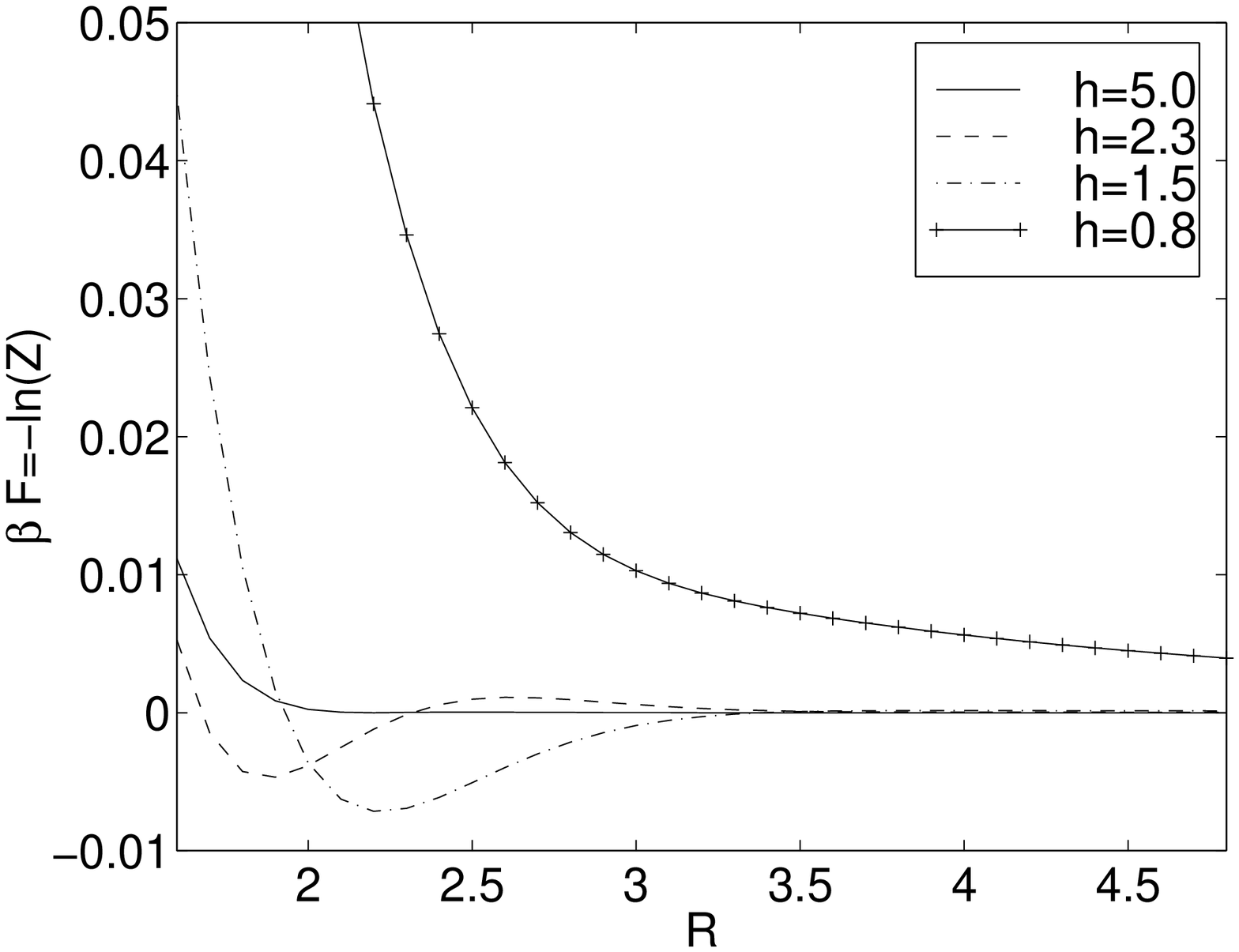}{Numerical results for the free energy vs.
separation distance $R$
of singly charged colloids near a single plasma edge; case I: $a=0.8$.}
\myfigure{6cm}{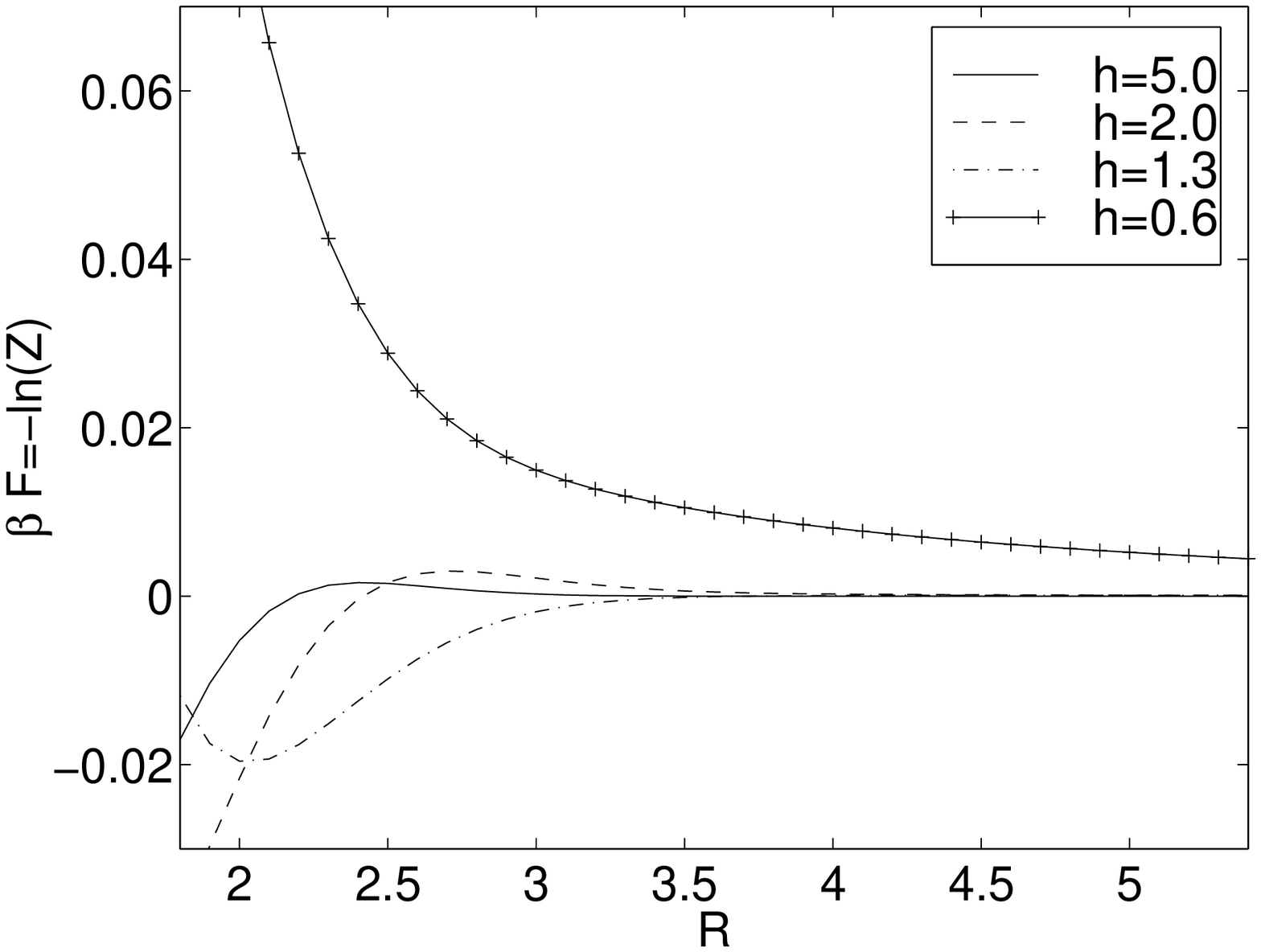}{Numerical results for the free energy vs.
separation distance $R$
of singly charged colloids near a single plasma edge; case II: $a=0.9$.}
\myfigure{6cm}{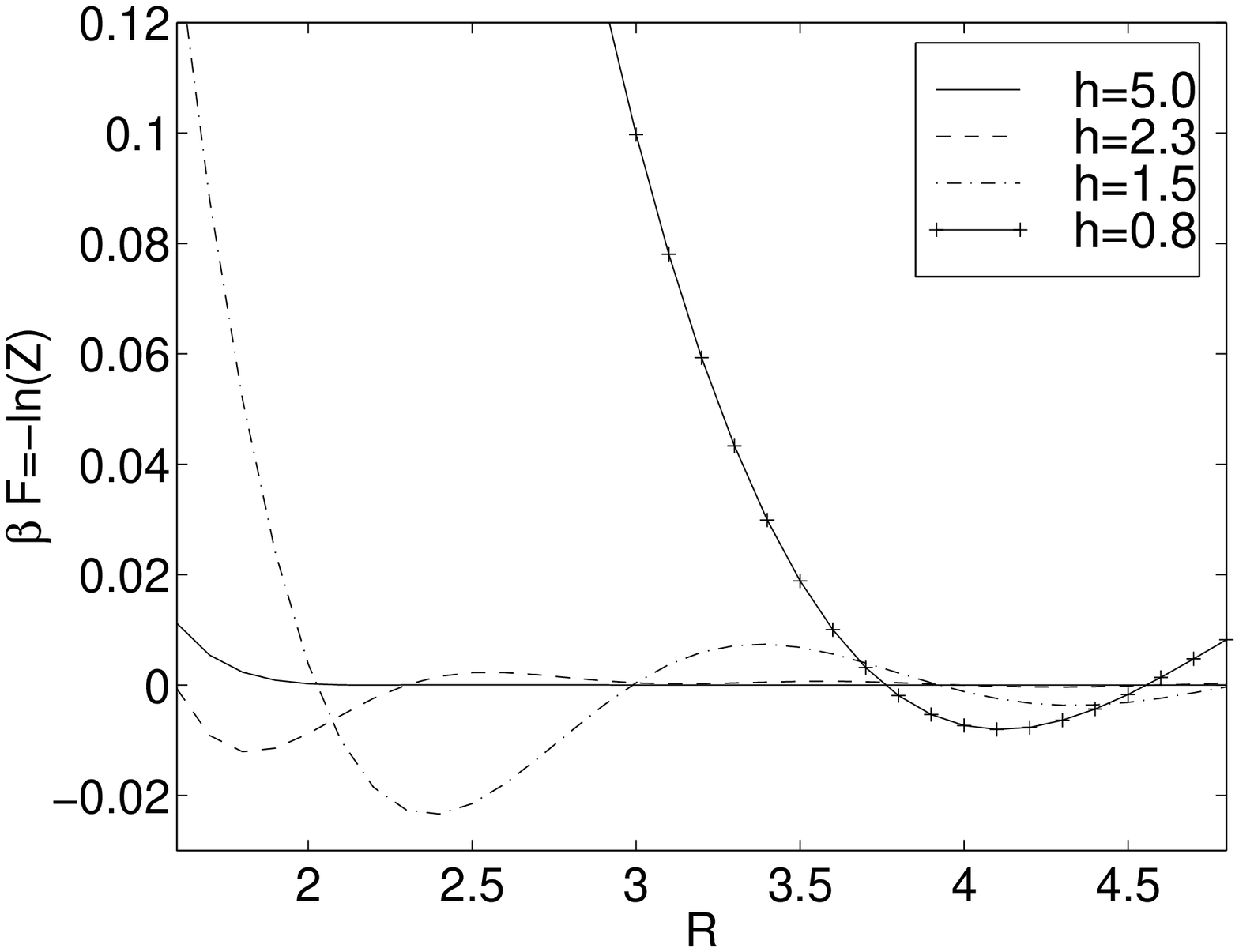}{Numerical results for the free energy vs.
separation distance $R$
of singly charged colloids near double plasma edges; $a=0.8$.}
\myfigure{6cm}{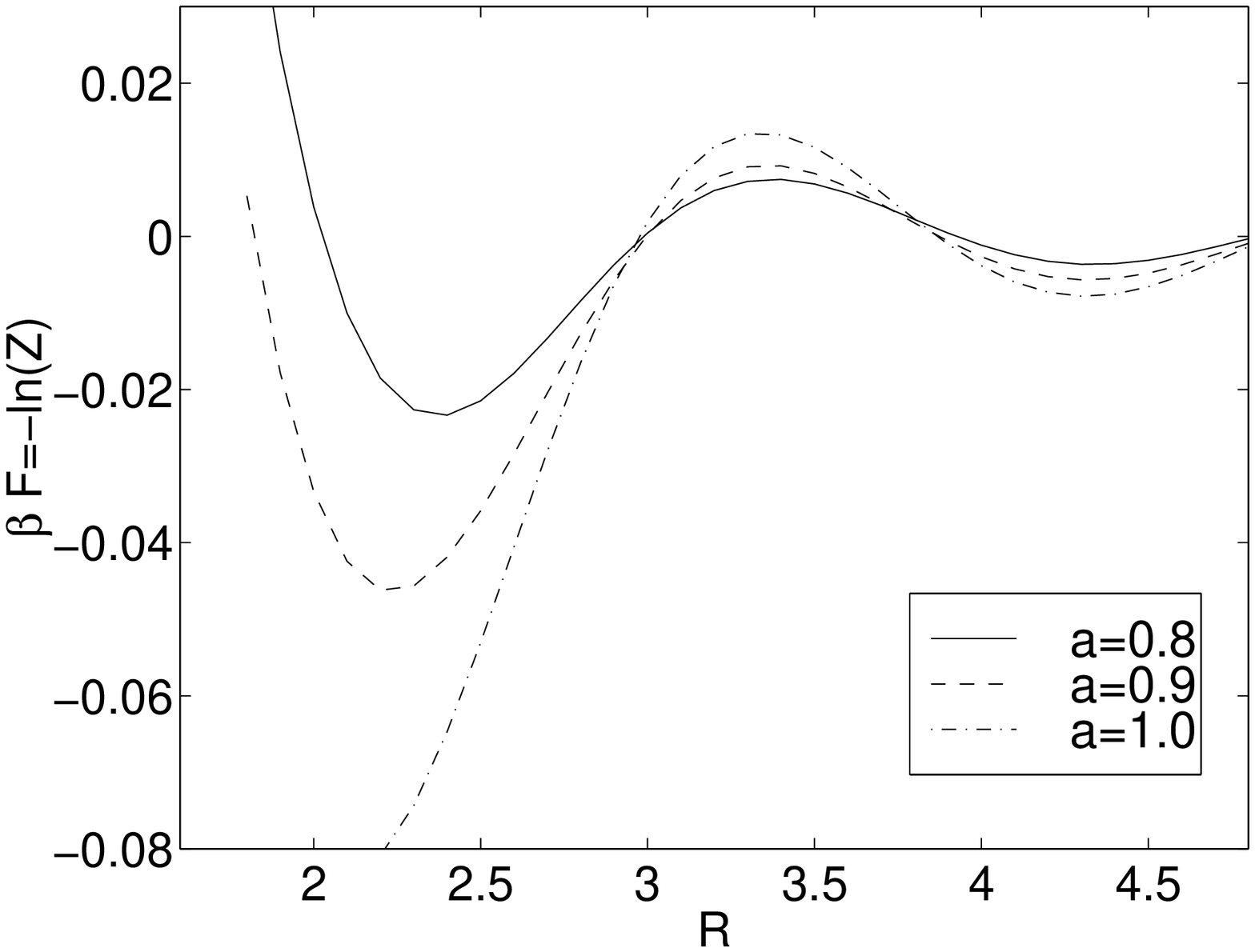}{Numerical results for the free energy vs.
separation distance $R$
of singly charged colloids near double plasma edges; $h=1.5$.}
\myfigure{6cm}{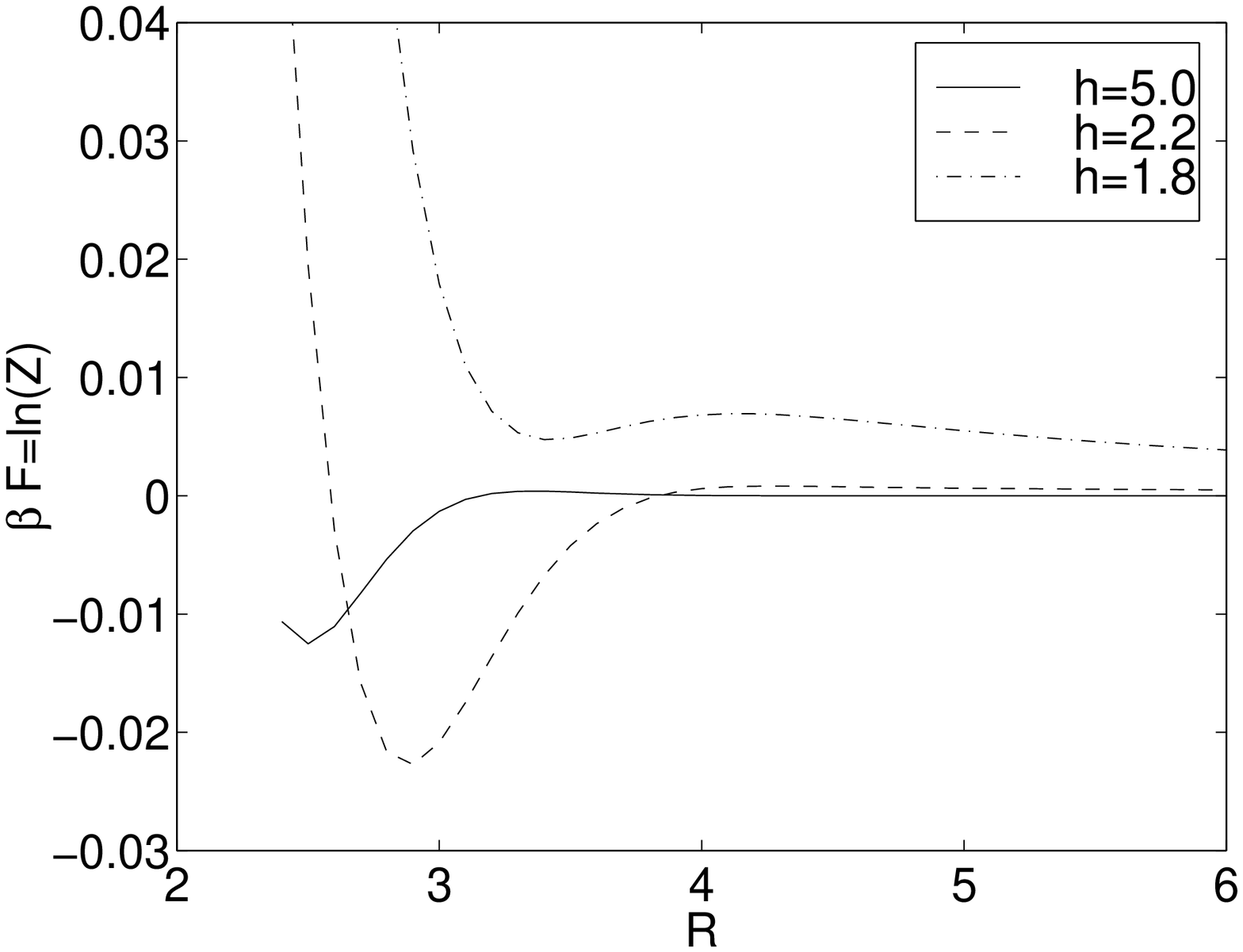}{Numerical results for the free energy vs.
separation distance $R$
of doubly charged colloids near a single plasma edge; $a=1.2$.}
\myfigure{6cm}{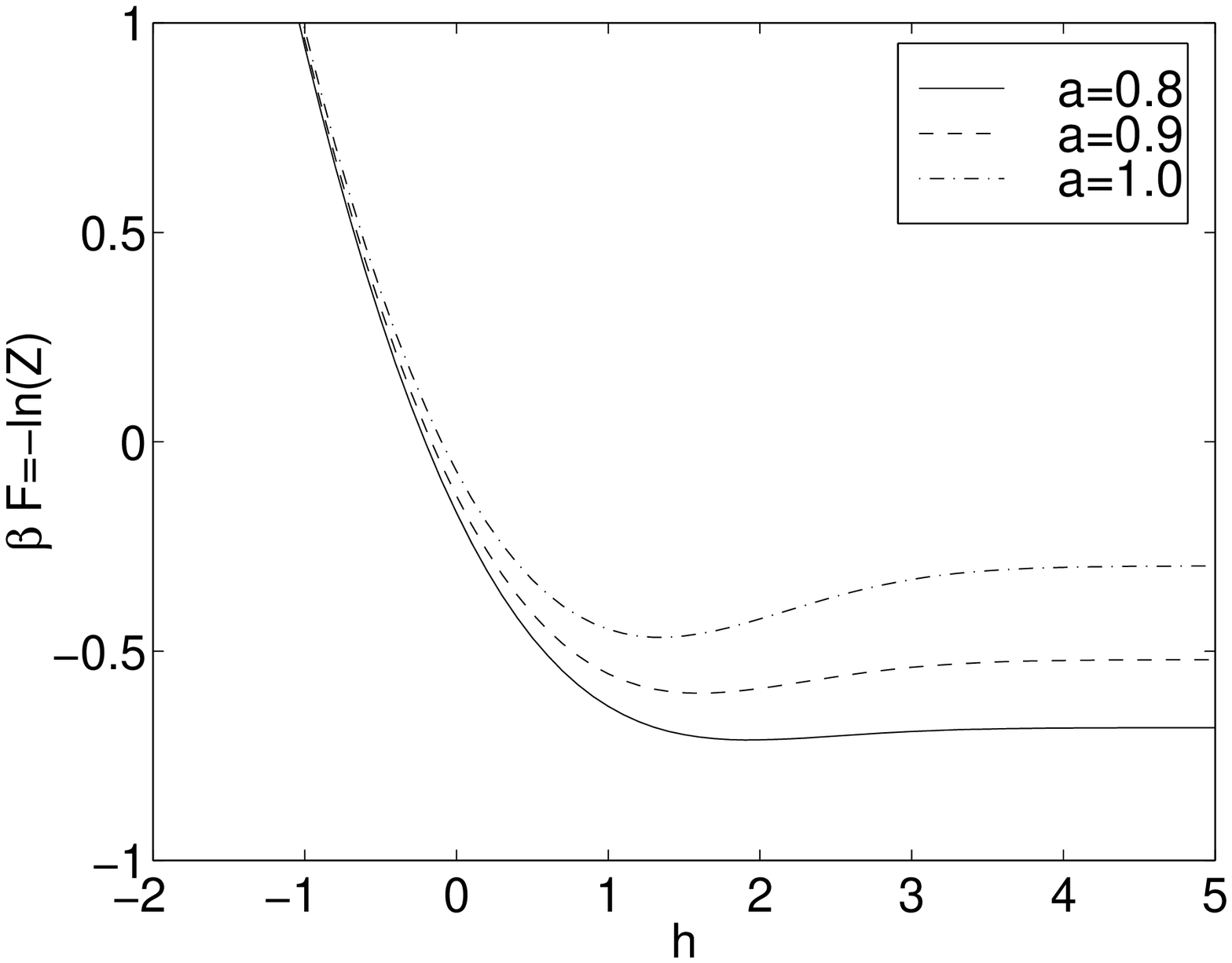}{Numerical results for the free energy for
one singly charged colloid near a single plasma edge.}

\begin{thebibliography}{99}

\bibitem{Kepler} 
G. M. Kepler and S. Fraden, \prl {\bf 73}, 356 (1994).

\bibitem{Grier} 
J. C. Crocker and D. G. Grier, \prl {\bf 77}, 1897 (1996);
A. M. Larsen and D. G. Grier, {Nature} {\bf 385}, 230 (1997);
D. G. Grier, {Nature} {\bf 393}, 621 (1998). 

\bibitem{Parsegian} 
R. Podgornik and V. A. Parsegian, \prl {\bf 80}, 1560 (1998).

\bibitem{Schmitz} 
K. S. Schmitz, {Langmuir} {\bf 13}, 5849 (1998).

\bibitem{Allahyarov} 
E. Allahyarov, I. D'Amico and H. L\" owen, \prl {\bf 81}, (1998).

\bibitem{Neu} 
J. C. Neu, \prl {\bf 82}, 1072 (1999).

\bibitem{Raimbault} 
E. Trizac and J. L. Raimbault, (eprint: cond-mat/9909420).

\bibitem{Sader} 
J. E. Sader and D. Y. Chan, 
{J. Colloid Interface Sci.} {\bf 213}, 268 (1999).

\bibitem{bren}
T. M. Squires and M. P. Brenner, (eprint: cond-mat/0003195);
E. R. Dufresne, T. M. Squires, M. P. Brenner and D. G. Grier, 
(eprint: cond-mat/0003314).

\bibitem{Jaco} 
B. Jancovici, \prl {\bf 46}, 386 (1981);
{J. Stat. Phys.} {\bf 28}, 43 (1982);
{J. Stat. Phys.} {\bf 29}, 263 (1982).

\bibitem{rich}
R. P. Sear, (eprint: cond-mat/0002249).

\bibitem{orland}
R. R. Netz and H. Orland, 
{European Phys. J. E} {\bf 1}, 67 (2000);
{European Phys. J. E} {\bf 1}, 203 (2000);
{European Phys. J. D} {\bf 8}, 145 (2000);
{Europhysics Lett.} {\bf 45}, 726 (1999);
\pre {\bf 60} 3174 (1999);
(eprint: cond-mat/9902220, cond-mat/9902085,  cond-mat/9807009).

\bibitem{brown}
L. S. Brown and L. G. Yaffe,
(eprint: physics/9911055).

\bibitem{ajl}
B. -Y. Ha and A. J. Liu, {\it Physical Questions Posed by DNA
Condensation}, to appear in {\it Physical Chemistry of
Polyelectrolytes}, ed. T. Radeva (Marcel Dekker, New York,
2000), (eprint: cond-mat/0003162);
\pre {\bf 58} 6281 (1998);
\pre {\bf 60} 803 (1999).

\bibitem{smg}
S. M. Girvin, 
{\it The Quantum Hall Effect: Novel Excitations and Broken Symmetries},
Les Houches Lectures, (Springer-Verlag, Berlin and Les Editions Physique, 
Paris, 2000), (eprint: cond-mat/9907002);

\bibitem{smg2}
S. M. Girvin and T. Jach, \prb {\bf 29}, 5617 (1984).

\bibitem{notes}
In a matrix notation, eq.(\ref{v1}) can be written as
\begin{eqnarray}
	 \O=1-[\Tm(w)\G\Tm^\dagger(w)+
	 	\Tm(\overline{w})\G\Tm^\dagger(\overline{w})].
	\nonumber
\end{eqnarray}
where, $G_{ij}\equiv\delta_{ij}\g{i}$.
Using this, the free energy is given by \begin{eqnarray}
	 \beta F\eq -\ln\det\O\nonumber\\
	 \eq -{\mathrm tr}\ln\{1-[\Tm(w)\G\Tm^\dagger(w)+
	 	\Tm(\overline{w})\G\Tm^\dagger(\overline{w})]\}\nonumber\\
	\eq\sum_{k=1}^\infty {1\over k}{\mathrm tr}
	 	[\Tm(w)\G\Tm^\dagger(w)+
	 	\Tm(\overline{w})\G\Tm^\dagger(\overline{w})]^k\nonumber\\
	\eq\sum_{k=1}^\infty {2\over k}{\mathrm tr}\G^k+F(w)\nonumber\\
	\eq 2\beta F_0 +\beta F(w)\nonumber
\end{eqnarray}
where, $\beta F_0\equiv -\ln\det(1-\G)$ represents the free energy of
each individual void, and $F(w)$ can be viewed as the
interaction potential.  

\end{thebibliography}
\end{document}